\documentclass[aps,prb,twocolumn,superscriptaddress,showpacs,notitlepage,reprint]{revtex4-1}
\pdfoutput=1
\usepackage{amsmath}
\usepackage{appendix}
\usepackage{bm}
\usepackage{graphicx}
\usepackage{txfonts}
\usepackage{color}

\newcommand{\bea}{\begin{eqnarray}}
\newcommand{\eea}{\end{eqnarray}}
\newcommand{\bei}{\begin{itemize}}
\newcommand{\eei}{\end{itemize}}
\newcommand{\be}{\begin{equation}}
\newcommand{\ee}{\end{equation}}
\newcommand{\bse}{\begin{subequations}}
\newcommand{\ese}{\end{subequations}}
\newcommand{\bfg}{\begin{figure}}
\newcommand{\efg}{\end{figure}}

\newcommand{\smeq}{\! = \!}

\newcommand{\ve}{\varepsilon}

\newcommand{\bese}{\begin{subequations}}
\newcommand{\eese}{\end{subequations}}

\begin{document}
\title{Spin-orbit induced chirality of Andreev states in Josephson junctions}
\author{Andres A. Reynoso}
\affiliation{ARC Centre for Engineered Quantum Systems, School of Physics, The University of Sydney, NSW 2006, Australia}
\author{Gonzalo Usaj}
\affiliation{Centro At{\'o}mico Bariloche and Instituto Balseiro,
Comisi\'on Nacional de Energ\'{\i}a At\'omica (CNEA), 8400 Bariloche, Argentina}
\affiliation{Consejo Nacional de Investigaciones Cient\'{\i}ficas y T\'ecnicas (CONICET), Argentina}
\author{C. A. Balseiro}
\affiliation{Centro At{\'o}mico Bariloche and Instituto Balseiro,
Comisi\'on Nacional de Energ\'{\i}a At\'omica (CNEA), 8400 Bariloche, Argentina}
\affiliation{Consejo Nacional de Investigaciones Cient\'{\i}ficas y T\'ecnicas (CONICET), Argentina}
\author{D. Feinberg}
\affiliation{Institut N\'eel, CNRS and Universit\'e Joseph Fourier, 38042 Grenoble, France}
\author{M. Avignon}
\affiliation{Institut N\'eel, CNRS and Universit\'e Joseph Fourier, 38042 Grenoble, France}
\date{\today}
\begin{abstract}
We study Josephson junctions (JJs) in which the region between the two
superconductors is a multichannel system with Rashba spin-orbit coupling
(SOC) where a barrier or a quantum point contact (QPC) is present. These systems might
 present unconventional Josephson effects such as Josephson currents
for zero phase difference or  critical currents that
\textit{depend on} the current direction. Here, we discuss how the
spin polarizing properties of the system in the normal  state affect the spin characteristic of the Andreev bound states inside the junction.
This results in a strong correlation between the spin of the Andreev states and the direction in which they transport Cooper pairs. While the current-phase relation for the JJ at zero magnetic field is qualitatively unchanged by SOC, in the
presence of a weak magnetic field a strongly anisotropic behavior and the
mentioned anomalous Josephson effects follow. We show
that the situation is not restricted to barriers based on constrictions such as
QPCs and should generically arise if in the normal system the direction of the
carrier's spin is linked to its direction of motion.
\end{abstract}

\maketitle

\section{Introduction}

The study of spin-orbit coupling (SOC) and related effects is one of the
most active fields in mesoscopic physics. As the fabrication techniques
improve, the link between the orbital and the spin degrees of freedom can be
engineered allowing for the exploration of quantum phenomena at a deeper
level.\cite{Winkler_book,spintronicsRMP,Spintronicsbook,AwschalomSPhysics09}
Moreover, the importance of SOC goes far beyond mesoscopic device design as,
for example, it has led to the identification of  the Quantum Spin Hall phase in topological insulators,\cite
{topo2007science} a new phase of matter.\cite
{KaneMele2005,topoinsulators2011rmp} More
recently, hybrid systems involving SOC and superconductivity have become one
of the most investigated platforms.\cite
{PhysicsAlicea2010,scienceHuntMajoranas2011} When in combination with an
appropriate magnetic field (generated externally or induced by proximity
with a magnet in the sample) the system is expected to develop topological
superconductivity.\cite
{FuKane2008,SauDasSarma2010,*Alicea2010prb,*OregRefaelvonOppen2010} This
opens the possibility to synthesize Majorana fermions in semiconducting quantum wires, cold atoms or devices based on topological insulators.

In this work we theoretically investigate the DC Josephson effect in a
hybrid system having superconductivity and spin-orbit coupling; more
precisely, two s-wave superconducting leads (S) coupled by a two-dimensional
normal region (N) that has Rashba spin-orbit coupling. For
superconductor-normal-superconductor (S-N-S) junctions, it has been
recognized that the interplay between external magnetic fields---or,
alternatively, intrinsic magnetism---and the spin-orbit coupling can change
the relation between the supercurrent and the phase difference between the
two superconductors.\cite
{Bezuglyi02,Krive04,Krive2005b,Dimitrova06,DellAnna07,BeriBB08,Zazunov2009,Malshukov2010,*Malshukov2011b,Ostaay2011edgeRashbaJosephson}

The key ingredient in the S-N-S junction studied here is that the normal
region \emph{contains} a gate-voltage controlled barrier, see, for instance,
the Josephson junction measured in Ref.~%
\onlinecite{Takayanagi95,*Takayanagi95prb}
In Ref.~\onlinecite{ReynosoUB08prl} a weak external magnetic field,
applied in a specific in-plane direction, was predicted to generate a
controllable phase shift in the current-phase relation (CPR), i.e., the
system develops a supercurrent for zero phase difference between the
superconductors. As the gate voltage is reduced, and the barrier allows the
transport of a few (more than one) transmission channels, the Josephson
current becomes non-symmetrical as a function of the phase. The maximum
dissipationless current, the critical current, depends on the direction of
the current and therefore there exists a current window on which the
supercurrent is rectified.

In the long Josephson junction limit with transparent S-N interfaces, a
normal region with a one-dimensional (1D) spin-orbit coupling combined with
a magnetic field, either parallel or perpendicular to the SOC axis, can only produce transitions between $0$-$\pi $ junctions
as it is the case for Superconductor-Ferromagnet-Superconductor (S-F-S)
junctions.\cite{Golubov04,Buzdin05revSF} In this limit this can be
understood using Kulik theory,\cite{Kulik69} but the same result follows by
virtue of symmetry relations of the Bogoliubov-de Gennes (BdG) Hamiltonian,
\cite{Liu2010,Margaris2012} irrespective of the geometry of the junction for
any generic 1D SOC coupling. On the other hand, the magnetic field triggers
shifts in the CPR by incorporating the mixing with other subbands in an
effective 1D model.\cite{Krive04,Krive2005b}

In the junction studied here the anomalous
effects arise because the barrier in the normal region behaves as an
unconventional spin polarizer.
As it was reported by Eto \textit{et al.} in Ref.~\onlinecite{EtoHK05}, a quantum point contact (or barrier) in a multichannel normal material
with Rashba SOC spin-polarizes the current without the need for magnetic
fields or ferromagnets.\cite
{SilvestrovM05,ReynosoUB06_Lawn,*ReynosoUB07,KrichH08,Ulloa2010qpc} The
larger the spin-orbit strength and the smoother the barrier, the more the
polarization grows. Due to time-reversal symmetry, the favored spin
projection reverses if the direction of the transport is inverted.

At zero magnetic field, we show how these gate-voltage controllable spin
polarizing features in the normal device induce a correlation between the
velocity\footnote{Andreev states are bounded states, however, through them Cooper pairs are transferred between superconductors. In this work we call \emph{velocity} of an Andreev state to the velocity of the associated Cooper pairs being transferred (see Eq.\eqref{EQ:j} and Fig.\ref{FG:f3}).} and the spin of Andreev states. As a result, the application of a magnetic field, along the direction in which the SOC polarizes the current,
triggers the mentioned anomalous Josephson effects.

We present a simple WKB picture
for the transport through the barrier that successfully explains the main
features of the exact numerical results for the Andreev states including
their spin texture. We also show that the anomalous Josephson effects are not
restricted to QPCs, they also appear in wider barriers (stripe-shaped
samples) for which the transverse modes are so close in energy that the
plateaus of quantized conductance can not be resolved when changing the gate
voltage. For both types of barriers, N-S junctions with reduced transparency
make the normal device behave as an Fabry-Perot interferometer;\cite
{ReynosoThesisDr} the total transmission and the polarization are maximized
at particular values of the gate voltage that controls the barrier. This
resonant behavior translates into stronger Josephson anomalies that can be
tuned by changing the gate voltage in a small amount.\cite{ReynosoUB08prl}

The paper is organized as follows, in Sec. \ref{SC:Sec2} we introduce the
model and we outline the polarizing mechanism of the normal device, in Sec.
\ref{SC:Results} we focus on how the spin properties of the Andreev states
result in the anomalous Josephson effects mentioned above, and conclusions are presented in Sec.\ref{SC:Conclusions}.

\section{The model and the properties of the normal device}
\label{SC:Sec2}
\subsection{Model for the junction}

\label{SC:Model} Our model describes a Josephson junction with a QPC or
barrier in a normal region that has spin-orbit coupling. The device,
illustrated at the top of Figure \ref{FG:f1}, includes two superconducting
leads---Left (${L}$) and Right (${R}$)---and a central normal part ($N$)
made out of a 2DEG in which the barrier is created with the help of
electrostatic gates. For convenience we assume that the three regions are
restricted to a stripe geometry, i.e. the potential becomes infinite outside
the region $0<y<W_y$. The Hamiltonian describing the leads is
\begin{eqnarray}
\hat{H}_{\gamma } &=&\int d\bm{r}\left\{\sum_\sigma\bm{\Psi }_{\sigma
}^{\dag }(\bm{r})\left(\frac{\bm{p}^{2}}{2m_s}-\mu \right)\bm{\Psi }_{\sigma
}^{}(\bm{r})\right.  \notag \\
&&\left.\vphantom{\frac{\bm{p}^{2}}{2m_s}}-\Delta _{\gamma }(\bm{r})\bm{\Psi
}_{\uparrow }^{\dag }(\bm{r})\bm{\Psi }_{\downarrow }^{\dag }(\bm{r})-\Delta
_{\gamma }^{\ast }(\bm{r})\bm{\Psi }_{\downarrow }^{}(\bm{r})\bm{\Psi }%
_{\uparrow }^{}(\bm{r})\right\}\,.  \label{EQ:Hleads}
\end{eqnarray}
Here $\gamma\! = \! {L},{R}$ and the integral is done in the $x<x_{L}$
semi-space or the $x>x_{R}$ semi-space for the left and right lead,
respectively. In the above expression, $m_s$ is the electron mass in the superconductor, $\mu $ is
the chemical potential and $\bm{\Psi}_{\sigma }^{\dag }(\bm{r})$ creates an
electron with spin $\sigma $ at position $\bm{r}\! = \!(x,y)$. We assume
that within the leads the superconducting order parameter is constant and
set $\Delta_{{L}}(\bm{r})=\Delta_0$ and $\Delta _{{R}}(\bm{r})=\Delta_0
\mathrm{e}^{i\phi }$. In the following we take $\Delta_0\! = \! 1.5$meV as
in Nb.

The normal part of the junction, $x_{L}<x<x_{R}$, is described by the
Hamiltonian of a 2DEG with Rashba SOC and the potential $V(\bm{r})$ that
defines the barrier,
\begin{eqnarray}
\hat{H}_{N} &=&\int d\bm{r} \sum_{\sigma,\sigma^{\prime}}\bm{\Psi }_{\sigma
}^{\dag }(\bm{r})\mathcal{H}_N \bm{\Psi }_{\sigma^{\prime}}(\bm{r}) ~,
\label{EQ:hc} \\
\mathcal{H}_N&=&\frac{1}{2m^{\ast }}\left(p_{x}^{2}+p_{y}^{2}\right)+\frac{%
\alpha }{\hbar}\left(p_{y}\sigma _{x}-p_{x}\sigma _{y}\right)+V(\bm{r}%
)-\mu\,,  \notag  \label{EQ:HnormalCont}
\end{eqnarray}
where $m^{\ast }$ is the effective mass of the 2DEG carriers, $\alpha $ is
the strength of the Rashba coupling and $\sigma_{i}$ is the $i$-component of
the spin operator. The potential $V(\bm{r})$ includes a constant term that
shifts the bottom of the conduction band of the 2DEG with respect to the one
in the superconductor.

As in Ref.~\onlinecite{ReynosoUB08prl} we consider thin superconducting
films parallel to the $(x,y)$-plane. In this configuration in-plane magnetic
fields lead only to Zeeman interaction. Therefore, the contribution to the
Hamiltonian due to an external in-plane magnetic field is
\begin{equation}
\hat{H}_{Z} =\int d\bm{r} \sum_{\sigma,\sigma^{\prime}}\bm{\Psi }_{\sigma
}^{\dag }(\bm{r}) \frac{g(x)\mu_B}{2}\left(B_{x}\sigma _{x}+B_{y}\sigma
_{y}\right) \bm{\Psi }_{\sigma^{\prime}}(\bm{r})~,
\end{equation}
where $B_{i}$ are the components of the magnetic field and the gyromagnetic
factor
\begin{equation}
g(x)=
\begin{cases}
g_N, & \text{for~}x_L<x<x_R \\
g_S, & \text{otherwise}
\end{cases}
\end{equation}
is different for the leads and the normal region. In the regime of parameters studied in this paper
the superconducting contacts are almost unaffected by the magnetic field.
Namely, we take the Zeeman energy $E_Z$ on the normal region, always smaller
than $20\%$ of the superconducting gap $\Delta_0$ in the S contacts. As the g-factor in the In-based 2DEG is at
least $5$ times larger that the g-factor in the superconductor, the Zeeman
energy in the superconductor is at most $\Delta_0/25$.

In the present work we solve the full Hamiltonian, $\hat{H} \! = \! \hat{H}%
_{Z}\! + \! \hat{H}_{N}\! + \! \hat{H}_{{L}}\! + \! \hat{H}_{{R}}$,
numerically using a finite difference approach.\cite{Cuevas96} This is done
by writing a tight-binding version of the Hamiltonian and using a recursive
Green function method.\cite{PastawskiMedina2001} The details of the approach are presented in Appendix
\ref{AP:A}. For $\Delta_0\! \neq \! 0$ one must work in a Nambu space
representation since $\hat{H}_{tot}$ is indeed a BdG Hamiltonian.\cite
{deGennesBook}

In the lattice model the tunneling between the normal and
superconducting materials is described by hopping Hamiltonians, $\hat{H}_{N,L}$ and $\hat{H}_{N,R}$. The sites at the interfaces are connected by a spin-conserving matrix element, $t_{b}$, that arises due to the kinetic term in the original Hamiltonian ($\propto p_x^2$). As shown in Eq.\eqref{EQ:APtunn}, there are also spin-flipping tunneling elements between those sites. These arise due to the symmetrization made to assure Hermiticity of the SOC term:\cite{Matsuyama2002}
\be
 -\frac{\alpha(x)}{\hbar} p_x \sigma_y \rightarrow -\frac{1}{2\hbar}\left(\alpha(x)p_x+p_x \alpha(x)\right)\sigma_y.
\ee
Clearly, such symmetrization is relevant only where $\partial_x\alpha(x)\neq 0$: here the interfaces between the normal region and the superconductors.

As assumed in the classic
Blonder-Tinkham-Klapwijk (BTK) approach for studying Andreev reflection,\cite
{BlonderTK82} non-ideal transparencies can be introduced through a sharp potential at the N-S interfaces in the continuous model. In the lattice model, the transparency of the barrier is controlled by the parameter $A$ such that
\begin{equation}
t_b = A\, \frac{t_N+t_S}{2},  \label{EQ:tNtS_A}
\end{equation}
where $t_S$ and $t_N$ are the hopping parameters for the superconducting and
normal regions, respectively. Notice that for realistic junctions $t_S$ and $%
t_N$ are different since they are chosen to simulate the Fermi velocity and
electron mass mismatches between the 2DEG and the superconductor. A
transparent junction corresponds to $A\! = \! 1$ since in such a case $t_b$
is the average of the hopping in the two regions minimizing the scattering
at the S-N interfaces; choosing $A<1$ allows for the
simulation of poorer transparencies. To achieve the full transparent
junction case ($Z\! = \! 0$ case of BTK) it is not enough to set $A\! = \! 1$
but we also need to assume that $t_S\! = \! t_N$. Besides the interface
parameter $A$, the junction is characterized geometrically by its total
length, ${L_N}\! = \! x_{R}-x_{L}$, by the properties of the barrier (see
Appendix \ref{AP:A}) and by the number of transverse channels in the normal
region.

\subsection{Normal device: spin polarization due to a barrier potential}

\label{SC:normal} Understanding normal transport through smooth barriers, in
presence of spin-orbit coupling, is a prerequisite to the investigation of
the JJs we treat here. Therefore, in this section, we focus on the case in
which the leads are not superconducting but instead they are: $(I)$ a
continuation of the normal region, or $(II)$ a metallic material $M$,
different from the spin-orbit coupled 2DEG in the central region. For the
case $(I)$ Eto \textit{et al.}\cite{EtoHK05} have shown
that a 2DEG with a QPC is able to spin polarize the current without the help
of an external Zeeman field. They assume a Rashba SOC, and study the ballistic transport through QPCs described by a quite general
saddle-like potential.

\subsubsection{Origin of the current polarization: a qualitative picture}

We focus on case $(I)$ for which the interfaces 2DEG-leads do not introduce
backscattering. The main underlying mechanism can be exemplified using two
conducting channels $n\! = \! 1,2$ in a Wentzel-Kramers-Brillouin (WKB)
picture.\cite{MerzbacherBook} At every position $x_a$ one obtains the
dispersion relation that would follow if the system were infinitely long
with the $y-$dependent potential, $\widetilde{V}_{x_a}(y)\! \equiv \!
V(x_a,y)$. We first exclude from the Hamiltonian of the normal region in Eq.(%
\ref{EQ:hc}) the part of the Rashba SOC proportional to the perpendicular
momentum, $\mathcal{H}^{\mathrm{so}}_\perp\! = \!\frac{\alpha}{\hbar}%
p_{y}\sigma_{x}$. For $\mathcal{H}^\parallel \! \equiv \! \frac{\mathbf{p}^2%
}{2m^*}-\frac{\alpha}{\hbar}p_{x}\sigma_{y} + \widetilde{V}_{x_a}(y)$, solutions at $x_a$ are uniquely identified by $k_x$, by the spin projection along the $y$-axis, $\sigma_y\! = \!\pm$, and by the transverse mode number $n$. For each transmission channel, $(n,\sigma _{y})$, the $k_x$ solution at position $x_a$ is $|\Phi_{x_a,n}\rangle|k_x%
\rangle|\sigma_y\rangle$, with eigenenergies,
\begin{equation}
E^{\parallel}_{(n,\sigma_y)}(x_a)=\frac{\hbar}{2m^*}(k_x -
k_\alpha\sigma_y)^2 - E_\alpha +E^V_{n}(x_a)~,  \label{EQ:disp}
\end{equation}
where $k_\alpha \! \equiv \! \frac{m^*\alpha}{\hbar^2}$, $E_\alpha\! \equiv
\! \frac{\hbar^2 k_\alpha^2}{2 m^*}\! = \! \frac{\alpha^2 m^*}{2 \hbar^2}$
and $E^V_{n}(x_a)$ are the transverse mode energies, namely, the
eigenenergies corresponding to the eigenfunctions, $\langle
y|\Phi_{x_a,n}\rangle$, of the Hamiltonian $\frac{p_y^2}{2 m^*}\! + \!
\widetilde{V}_{x_a}(y)$.

\begin{figure*}[t]
\centering
\includegraphics[width=.93\textwidth]{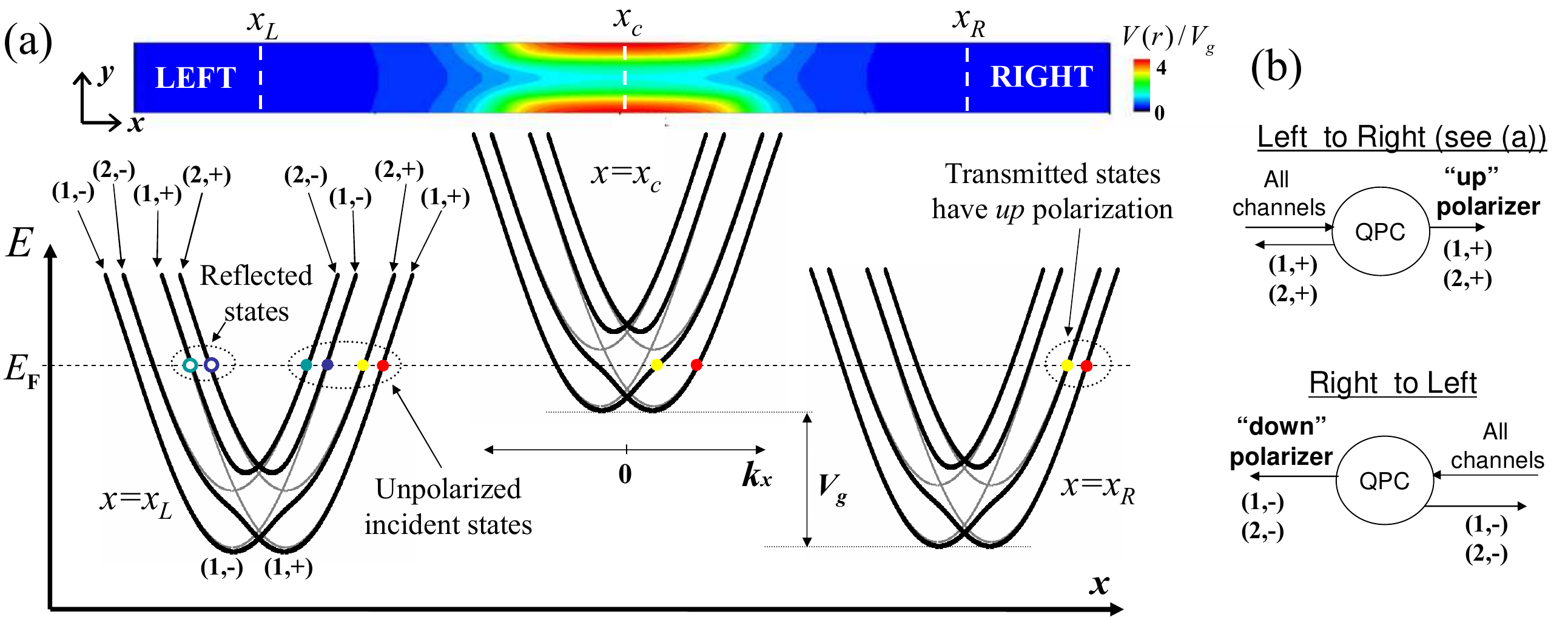} \vspace{-0.3cm}
\caption{(Color online) (a) UPPER PANEL: scheme of the junction including
the leads and a qualitative color plot of the barrier potential, $V({\bm r})$%
; $V_g$ is the value at the saddle-point. LOWER PANEL: qualitative two-mode
picture of how the current through the barrier becomes spin polarized (left
to right case). In the dispersion relations $E(k_x)$ at different positions
we indicate the relevant states using dots at the Fermi energy, $E_\mathrm{F}
$. Electrons from the Left lead (at $x\! = \! x_{L}$) are injected in four
states, $(n, \sigma_y)$, with $n\! = \! 1,2$ and $ \sigma_y\!
= \!\pm$. The barrier allows only one transverse mode to be transmitted
(first conductance plateau, $G=2e^2/h$). Electrons injected at $x\! = \!
x_{L}$ in states $(1,+)$ and $(2,+)$ are transmitted to the right. They
arrive at $x\! = \! x_{R}$ in states $(1,+)$ and $(2,+)$ and therefore the
current is $up$ polarized along $y$-direction. Notice that---in the passage
through the barrier---an electron injected in state $(1,+)$ does not see any
avoided crossing (as in a zero-SOC barrier) whereas an electron injected in
state $(2,+)$ suffers two SOC-induced avoided crossings: first it is mixed
with $(1,-)$ and then it returns to $(2,+)$. (b) Summary of the left to
right and right to left transport properties of the barrier. The spin
polarization from right to left is $down$ as electrons injected (at $x\! =
\! x_{R}$ from the Right lead) in $(1,-)$ and $(2,-)$ are transmitted
whereas electrons injected in $(1,+)$ and (2,+) are reflected to $(2,-)$ and
$(1,-)$, respectively.}\label{FG:f1}
\vspace{-0.5cm}
\end{figure*}

Importantly, the $E_{(n,\sigma
_{y})}^{\parallel }(x_{a})$ parabolas for modes $(1,-)$ and $(2,+)$ (as well
as for $(1,+)$ and $(2,-)$) cross each other; see gray lines in Fig.~\ref
{FG:f1}(a). Away from the center of the barrier and for strong enough Rashba
interaction these crossings are \emph{below} the Fermi level, $E_{\mathrm{F}%
} $. The transverse SOC, $\mathcal{H}^{\mathrm{so}}_\perp \propto p_y \sigma_x$, produces avoided crossings because it coherently mixes transverse
modes with different parity, $n\! \mod 2$, (since $p_{y}=-\mathrm{i}\hbar \frac{%
\partial }{\partial y}$) and opposite $\sigma _{y}$. Figure \ref{FG:f1}(a)
sketches the transport through the QPC in the adiabatic picture. For the
situation shown there, the value of the potential at the saddle point, $V_{g}$, is
tuned to allow the passage of a single transverse mode: the conductance is $%
2e^{2}/h$. As electrons at $E_{\mathrm{F}}$ pass through the QPC from left
to right their $k_{x}$ momentum change. Electrons in the lowest-lying split
channel $(1,+)$ pass through unaffected. On the other hand, electrons in
the higher split channel $(2,+)$ are swept twice through the anticrossing:
first changing to $(1,-)$ and then to $(2,+)$. Incoming electrons occupying the \emph{down} modes $(1,-)$ and $(2,-)$ (see contact with the Left lead at position $x_{L}$) do not arrive at $x_{c}$ (center of the barrier) but
instead they are backscattered to $x_{L}$ into the left-traveling modes $%
(2,+)$ and $(1,+)$, respectively. Thus, the current arriving at
$x_{R}$ (the position of the contact with the Right lead) is \emph{up}-spin
polarized along the $y-$direction.

In the two-modes sketch of Fig.~\ref{FG:f1} we have assumed that the avoided
crossings are fully effective. In a more general case the two quantum
channels associated to the first transverse mode can be divided in: (i)
channel $(1,+)$ which is fully transmitted to $x_{R}$ without state
conversion, and (ii) combination of $(2,+)$ and $(1,-)$ that arrives to $%
x_{R}$ after mixing in region of the barrier. The larger the strength of the
Rashba coupling and/or the smoother the barrier potential, the more
effective are the avoided crossings and the better becomes the polarizer:
the contribution to the current of the $(2,+)$ quantum channel is increased.
As the polarization is defined by the transport properties of electrons at
the Fermi level, what matters most is the gradient of the barrier potential
at the positions $x^{\mathrm{ac}}$ in which the crossings between the relevant dispersions fall at $E_\mathrm{F}$, i.e.,
\begin{equation}
E_\mathrm{F}=E^{\parallel}_{(1,\sigma_y)}(x^{\mathrm{ac}})=E^{\parallel}_{(2,%
\bar{\sigma}_y)}(x^{\mathrm{ac}})~.
\end{equation}
Two positions, $x^{\mathrm{ac}}_{{L}}<x_c$ and $x^{\mathrm{ac}}_{{R}}>x_c$, fulfill this condition. If the barrier is symmetric with respect to $x_c$
both avoided crossings share the same effectiveness. In Ref.~%
\onlinecite{EtoHK05} the probability of state conversion is
estimated---using the Landau-Zener formula for the case that $\widetilde{V}%
_{x}(y)$ is a hard-wall constriction potential of variable width $W(x)$---as
$p_{{L},{R}}\! = \! 1- \mathrm{e}^{-2\pi\lambda}$ with $\lambda\! \approx \!
k_{\alpha} \left|W(x^{\mathrm{ac}}_{{R}})/[\frac{dW}{dx}(x^{\mathrm{ac}}_{{L}%
,{R}})]\right|$.

Notice that for barriers with coexistence of Rashba and Dresselhaus SOCs,
nonzero polarization arises in exactly the same way along the direction of
the spin operator that multiplies $p_\theta$, with $\hat{\theta}\! \equiv
\!(\cos\theta,\sin{\theta})$, the direction parallel to the current that
flows through the barrier (here $\hat{\theta}\! = \!\hat{x}$ and the
corresponding spin operator for the Rashba SOC is $\sigma_y$). This opens
the possibility to electrically change the spin-polarization axis of an Eto
\textit{et al}. polarizer as the Rashba strength is tuned with gate-voltage\cite{NittaATE97}
and the Dresselhaus interaction remains constant.
\vspace{-0.8cm}
\subsubsection{Time-reversal symmetry and transport from right to left}
In the above discussion we have focussed on the
left (${L}$) to right (${R}$) transport in the symmetric case. We now relax the condition of
spatial symmetry of the barrier profile allowing it to be different for $%
x<x_{c}$ and for $x>x_{c}$. Therefore, we assume that the probability to follow the avoided crossing is $p_{L}$ at $x_{{L}}^{\mathrm{ac}}$ and $p_{R}$ at $x_{{R}}^{%
\mathrm{ac}}$. For the two-channel model, with the barrier in the first plateau of conductance, the nonzero
transmission coefficients (from left to right) at the Fermi energy are
\bese \label{EQ:Tres}
\begin{eqnarray}
T_{({L},1,+)\rightarrow ({R},1,+)} &=&1~, \\
T_{({L},1,-)\rightarrow ({R},1,-)} &=&(1-p_{L})(1-p_{R})~, \\
T_{({L},1,-)\rightarrow ({R},2,+)} &=&(1-p_{L})p_{R}~, \\
T_{({L},2,+)\rightarrow ({R},1,-)} &=&p_{L}(1-p_{R})~, \\
T_{({L},2,+)\rightarrow ({R},2,+)} &=&p_{L}p_{R}~.
\end{eqnarray}
\eese
The conductance, $G$, the polarization of the transmitted electrons, $P$,
and the relative spin polarization of the incoming electrons that contribute
to the current, $D$, can be expressed in terms of the transmission
coefficients as,
\bese \label{EQ:defGPD}
\begin{eqnarray}
G_{{L}\rightarrow {R}} &=&\frac{e^{2}}{h}\sum_{n_{{L}},\sigma _{y,{L}%
}}\sum_{n_{R},\sigma _{y,{R}}}T_{({L},n_{L},\sigma _{y,{L}})\rightarrow ({R}%
,n_{R},\sigma _{y,{r}})}~, \\
P_{{L}\rightarrow {R}} &=&\frac{e^{2}}{hG}\sum_{n_{{L}},\sigma _{y,{L}%
}}\sum_{n_{R},\sigma _{y,{R}}}\sigma _{y,{R}}T_{({L},n_{L},\sigma _{y,{L}%
})\rightarrow ({R},n_{R},\sigma _{y,{R}})}~,~~~~~~~~~ \\
D_{{L}\rightarrow {R}} &=&\frac{e^{2}}{hG}\sum_{n_{{L}},\sigma _{y,{L}%
}}\sum_{n_{R},\sigma _{y,{R}}}\sigma _{y,{L}}T_{({L},n_{L},\sigma _{y,{L}%
})\rightarrow ({R},n_{R},\sigma _{y,{R}})}~,
\end{eqnarray}
\eese
where the subscript ${{L}\rightarrow {R}}$ refers to the left to right transport. For the polarizing barrier, we introduce the coefficients of Eq.\eqref{EQ:Tres} in Eq.\eqref{EQ:defGPD} obtaining: $G_{L\rightarrow R}\smeq\frac{%
2e^{2}}{h}$,~ $P_{L\rightarrow R}\smeq p_{R}$ and $D_{L\rightarrow R}\smeq p_{L}$.

Importantly, due to time-reversal symmetry the resolved transmission coefficients at the
Fermi energy fulfill the following relation
\begin{equation}
T_{(L,n_L,\sigma_L)\rightarrow (R,n_R,\sigma_R)}=T_{ (R,n_R,\bar{\sigma}%
_R)\rightarrow (L,n_L,\bar{\sigma}_L)}~.  \label{Eq:TresTRS}
\end{equation}
Therefore the resolved transmission in the opposite direction of electrons
with opposite spins (at both ends) is identical. This follows from the
self-duality of the scattering matrix in a spin-1/2 time-reversal symmetric
system.\cite{Dyson1962} Here, this expression is of great importance for the
Josephson effect because at the S-N interfaces an electron coming from the
normal region can be Andreev-reflected back to the normal as the lack of an
electron with opposite spin (a hole) in the same transverse mode. Since this
hole propagates as the lacking electron, Eq.\eqref{Eq:TresTRS} establishes
the main difference with a conventional magnetism-based polarizer. In the
latter systems, irrespective of the direction of the current, the largest
transmission is for electrons with spins parallel to the majority-spin.
\begin{figure}[b]
\centering
\includegraphics[width=.48\textwidth]{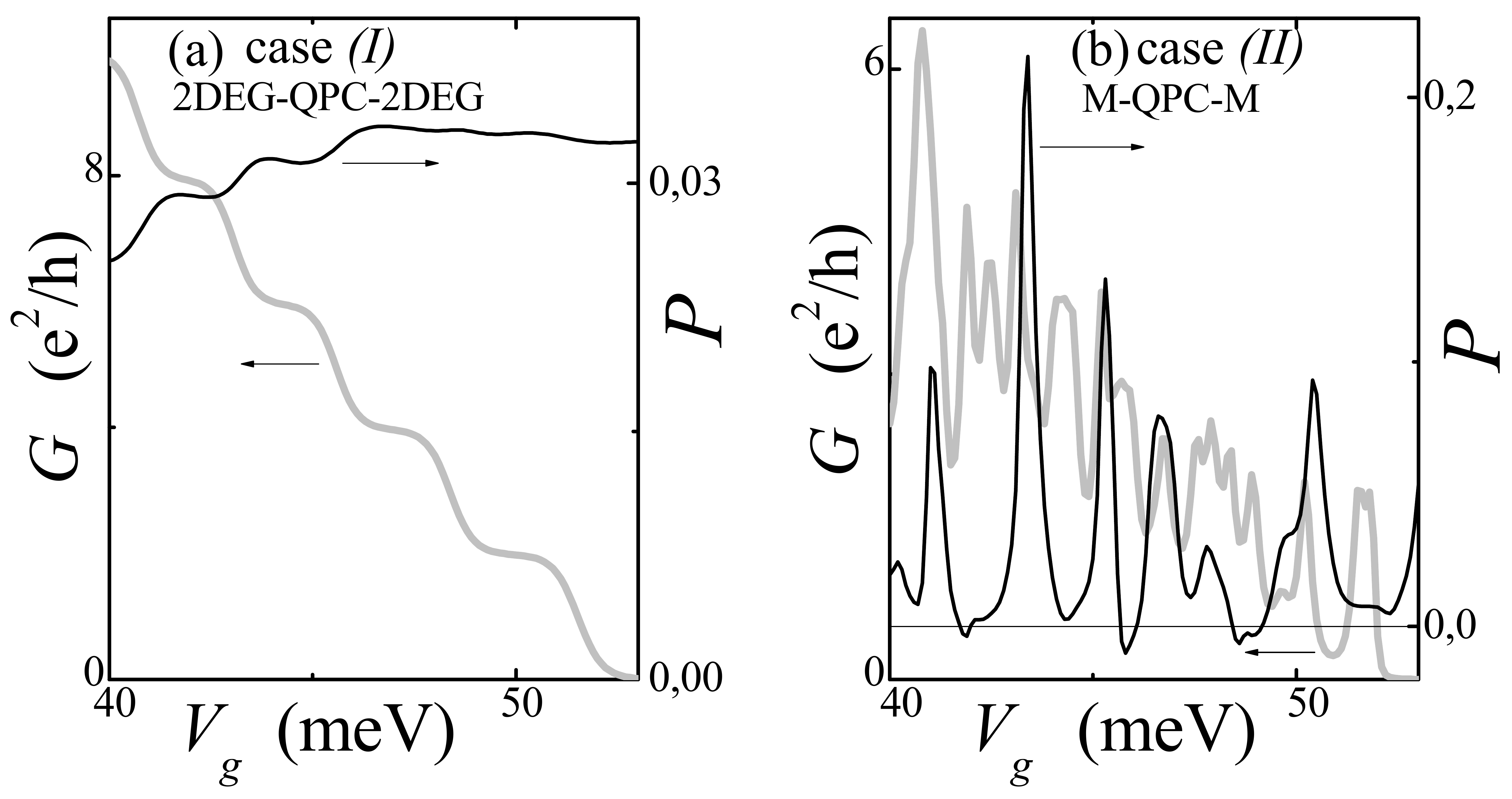} \vspace{-0.7cm}
\caption{(Color online) Conductance $G$ (gray line) and polarization $P$
(black line) as a function of $V_g$ for the quantum point contact, $\mathrm{%
QPC}_2$, described by Eq.\eqref{EQ:potential} taking $(z;W_b; L_b)\! =
\!(90; 100; 250)$nm. The 2DEG parameters are for an InAs-based 2DEG, ($n\! =
\!10^{12}\mathrm{cm}^{-2}$ and $m^*\! = \!0.045 m_e$ with $m_e$ the electron mass) $E_\mathrm{F}\! = \!
53.3$meV, $ \alpha\! = \! 5$meVnm and ${L_N}\! = \! x_R\! - \! x_L \!
= \! 1.2 \mu$m. In the simulation we take $W_y\! = \!180$nm and $%
a_0\! = \!3$nm. (a) Case $(I)$, 2DEG-QPC-2DEG: the leads are the same
2DEG and therefore the scattering at the interfaces is absent. As the Rashba
strength is weak the polarization is small: less than $4\%$ in the first
plateau. (b) Case $(II)$, M-QPC-M : The leads are a metallic material, $%
M $, different from the 2DEG leading to a strong backscattering at the
M-2DEG junctions; taking metallic Nb as a reference we use $m^*/m_s \! = \!
0.045$ and the Fermi velocities ratio, $v_\mathrm{F}^{\mathrm{2DEG}}/v_%
\mathrm{F}^{\mathrm{Nb}}\! \approx \!2.25$. Due to Fabry-Perot
interference-like effects both the conductance and the polarization show resonant
behavior as a function of $V_g$; remarkably, there are values of $V_g$ in
which the polarization can be much larger than the sub $4\%$ values observed
when the device is in between 2DEG leads.}\label{FG:f2}
\vspace{-0cm}
\end{figure}

For the polarizing barrier treated here, by combining the time reversal relation with
Eqs.\eqref{EQ:Tres} and \eqref{EQ:defGPD}, we get that the normal transport
from right to left has $G_{R\rightarrow L}=\frac{2e^2}{h}$, $P_{R\rightarrow
L}=-p_L$ and $D_{R\rightarrow L}=-p_R$. Note that for the case of the smooth
symmetrical barrier with strong Rashba of Fig.~\ref{FG:f1} $%
p_L=p_R\! \approx \! 1$ and thus $P_{R\rightarrow L}\! = \!-P_{L\rightarrow
R}\! \approx \! 1$. The inversion of the sign of the polarization is
summarized in Fig.~\ref{FG:f1}(b). It can also be understood from the WKB
picture in Fig.~\ref{FG:f1}(a) by noting that the electrons arriving to $L$ are due to incoming electrons from $R$ entering the sample: (i) in channel $(1,-)$ passing through the barrier without mixing or (ii) in channel $(2,-)$ arriving to $L$ in channel $(2,-)$ after mixing in the region of the barrier with channel $(1,+)$.

We have introduced the two most important differences between the Eto
\textit{et al}. polarizer and a magnetism-based spin polarizer. First, here
the device polarizes even though the conductance is equivalent to two
quantum channels, this contradicts the naive expectation that $G\! = \! 2 e^2/h$
implies spin degeneracy and that therefore the polarization should
be zero. The second distinctive feature is that in symmetrical barriers the
spin polarization of the transmitted current reverses when the current flow
is reversed. More generally, irrespective of the spatial symmetry of the
barrier, we have that $P_{L\rightarrow R}\! = \! -D_{R\rightarrow L}$ and $%
P_{R\rightarrow L}\! = \! -D_{L\rightarrow R}$ which means that the SOC
induced polarization does not conspire against the formation of Andreev
states. The preferred spin direction for the transmitted electrons arriving
to a given N-S interface is exactly opposite to the preferred spin direction
for the incoming electrons (or lack of them) that can be transmitted to the
other side of the barrier. In Sec.~\ref{SC:Results} we study which
properties of Andreev states are affected when these SOC-based polarizer
barriers are present inside a S-N-S Josephson junction and how these
properties manifest when a weak Zeeman energy is present.

\subsubsection{Numerical results in more realistic models}

\label{SC:normalRealistic} Using the QPC potential of Eq.\eqref{EQ:potential}
and the tight-binding model we compute the conductance and the polarization
in devices with many channels. In Fig.~\ref{FG:f2} results are presented as a function of $%
V_{g}$ for case $(I)$ and $(II)$ described above. We
have chosen the opposite limit to the one shown in Fig.~\ref{FG:f1}: the SOC
strength is weak, the avoided crossings are not fully effective and
thus the polarization is not strong. Case $(I)$ is shown in Fig. \ref
{FG:f2}(a), as the leads are 2DEG-based the scattering back to the barrier
is minimized and the plateaux of conductance are well defined. The
SOC-induced current polarization is always smaller than $5\%$, the
characteristic value in the first plateau is $P^{1st}_{(I)}\! = \! 0.04$.

Figure \ref{FG:f2}(b) shows results for the case $(II)$ in which the leads
are metals and therefore the electrons at the 2DEG-M interfaces are strongly
scattered back to the 2DEG region. Resonances in the conductance are
observed for special values of $V_{g}$. Correlatively, the spin polarization
of the current shows maxima too. Notice that the average spacing between
these maxima is larger than that of the conductance maxima. The main result
is that constructive interferences, due to multiple reflections of channels
between the interfaces and the QPC, strongly boost the current spin
polarization with respect to case $(I)$. We have checked (not shown) that
the relative polarization enhancement, $P^{\mathrm{peak}%
}_{(II)}/P^{1st}_{(I)}$, grows the weaker is the SOC strength. It is
interesting to analyze this amplification mechanism for the same phenomenon
is correlated with the anomalous Josephson current amplitude in the S-QPC-S
case.

A simplified scattering theory is sufficient for a qualitative
understanding.\cite{ReynosoThesisDr} It includes the two lowest channels,
the M-2DEG interfaces with a reflection coefficient $r_{0}$, the SOC
strength through the probability $p$ of state conversion at the avoided crossing
between channels $(1,\sigma_y)$ and $(2,\bar{\sigma}_y)$, and the phases
accumulated through the QPC for each of the quantum channels. These phases
can be calculated by a WKB formula, they depend both on the junction length
and on the gate voltage. The result of this calculation is that the full
transmission (thus the conductance) as well as the spin polarization
oscillate with the accumulated phases due to Fabry-Perot interferometry-like effects. As in the exact full numerical
simulation presented in Fig.~\ref{FG:f2} for a much larger total number of
channels, taking a large $r_{0}$ value one sees: (i) the ratio between the
polarization value at the peaks and the polarization for $r_{0}\! = \! 0$ is
larger the weaker is the Rashba strength; and (ii) the oscillation period
for the polarization is larger than the one of the conductance. The
explanation for this is that the spin polarization of the current crucially
depends on a coherent mixing of split channels with opposite spin
directions. The polarization is
controlled by the phase accumulated in the region \emph{between} the output
interface, at $x_R$, and the closest mixing location, at $x^{\mathrm{ac}}_R$%
. On the other hand, the full transmission is sensitive to the phase
acquired through the whole length of the device, thus explaining the faster
variation of the conductance with the gate voltage.\cite{ReynosoThesisDr}

\section{The effect of SOC-based polarizers in Josephson junctions}

\label{SC:Results}
Defining the superconducting phase difference, $\phi\! \equiv
\!\phi_L-\phi_R $, the Josephson current is given by,\cite
{Josephson62,deGennesBook}
\begin{equation}
I=\frac{2e}{\hbar }\left\langle \frac{\partial \hat{H}}{\partial \phi }%
\right\rangle = \frac{2e}{\hbar}\int_{-\infty}^{\infty}{J(\varepsilon,%
\phi)f_0(\varepsilon)\,d\varepsilon}~,  \label{EQ:iphi0}
\end{equation}
where $\left\langle ...\right\rangle $ indicates the expectation value at
thermal equilibrium , $f_0(x)\! = \! 1/(1\! + \! \mathrm{e}^{x/k_B T}) $ is
the Fermi distribution and we have introduced the Josephson current density
\begin{equation}
J(\varepsilon ,\phi )=\sum_{j}\frac{\partial \varepsilon _{j}}{\partial \phi
}\,\delta (\varepsilon -\varepsilon _{j}(\phi ))\,.  \label{EQ:j}
\end{equation}
Thus $J(\varepsilon ,\phi)$ contains the information about the velocity at which Cooper pairs (i.e., charge $2e$) are transmitted via Andreev bound
states, namely, the states with energy $\varepsilon_{j}$ which have nonzero $
\frac{1}{\hbar}\frac{\partial \varepsilon_{j}}{\partial \phi}$. As we discuss below, the
eigenvalues $\varepsilon_{j}$ of the system do not need to be computed in
order to get the Andreev levels. It is sufficient to compute $J(\varepsilon
,\phi)$ and later identify the coordinates $(\varepsilon ,\phi)$ at which $%
J(\varepsilon ,\phi)$ contributes to the current. In the discretized model
presented in Appendix \ref{AP:A} the current is obtained from Eq.%
\eqref{ECjj:iphi}. The calculation involves expectation values that can be
expressed as integrals over $\varepsilon$ of retarded and advanced Green
function.\cite{Zubarev60}. This method allows us to directly obtain $%
J(\varepsilon ,\phi)$ and from it identify the contributing Andreev energy
levels.

Notice that in the regime of parameters we work, $4\pi -$periodic components
are not present as the magnetic field is weak and it is applied \emph{%
parallel} to the direction of the SOC field; this is not the regime appropriate for Majorana physics.\cite{SauDasSarma2010,*Alicea2010prb,*OregRefaelvonOppen2010,PotterLee2010}

Here, due to the SOC-polarizing barriers and by virtue of a small magnetic
field along the polarization direction, we find strongly asymmetric ($2\pi $%
-periodic) $I(\phi)$ shapes. We have verified that the shapes always
fulfill
\begin{equation}
\int_{0}^{2\pi }I(\phi )d\phi =\frac{2e}{\hbar }\left( \langle \hat{H}%
\rangle _{\phi =2\pi }-\langle \hat{H}\rangle _{\phi =0}\right) =0\,,
\label{EQ:zeroArea}
\end{equation}
i.e., the area below the CPR is zero. This trivial property is not evident
from the plots reported in the following sections.

\subsection{Anomalous Josephson effects enhanced by \newline
barriers with SOC: ideal interfaces}

\label{SC:ResultsTR} In this section we study the case of ideal S-N
interfaces in order to isolate the effect in the Josephson current due to
the presence of SOC-induced spin polarization---in Sec.\ref{SC:nontransp} we
will focus on how the basic properties discussed here are modified by the
resonant-like behavior that appears for more realistic interfaces.\cite
{ReynosoUB08prl} The effective mass and Fermi velocity of the
superconducting leads are matched to the ones of the 2DEG and normal
reflection (for $|\varepsilon|<\Delta_0$) is minimized. In the lattice model
this is achieved by setting $t_S\! = \! t_N\! = \! t_b$ [$A\! = \! 1$ in Eq.%
\eqref{EQ:tNtS_A}].

\subsubsection{Spin properties of Andreev states in absence of magnetic field%
}

\begin{figure}[t]
\centering
\includegraphics[clip,width=0.48\textwidth]{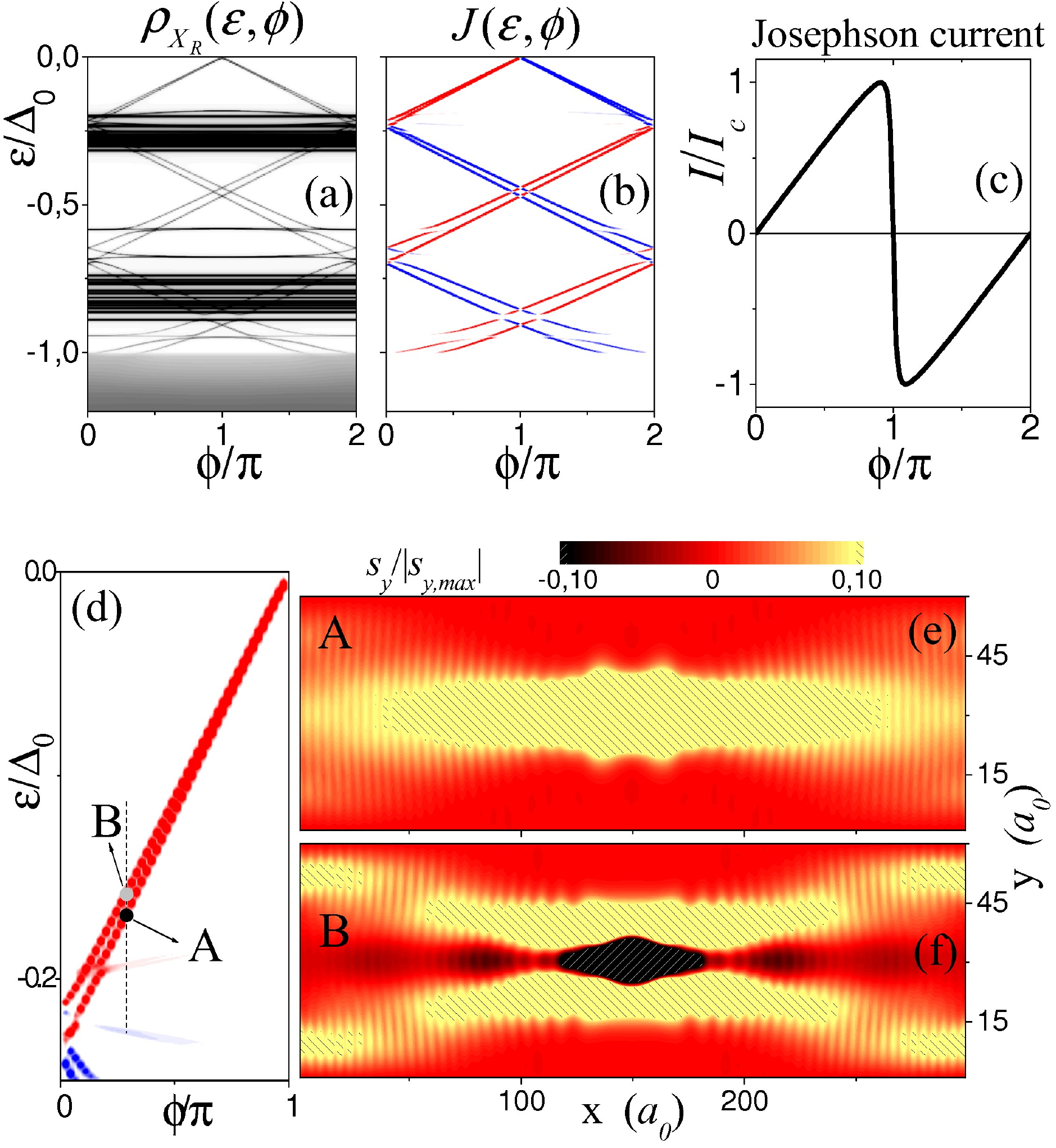} \vspace{-0.6cm} \vspace{%
-0.3cm}
\caption{(Color online) Exact results for the full multichannel model of a
Josephson junction with a SOC-based spin-polarizing barrier. The
superconductors have $\Delta _{0}\!=\!1.5$meV and are chosen to minimize
normal reflection (see text). The normal region length is ${L_{N}}%
\!=\!300a_{0}\!=\!900$nm and the QPC parameters are those given in Fig.~\ref
{FG:f2}. Here, the gate voltage is set in the middle of the first
conductance plateau and the SOC is strong, $ \alpha \!=\!40$meVnm, so
that the polarization is close to one. Color maps as a function of energy, $%
 \varepsilon $, and the superconducting phase difference $ %
\phi $ for: (a) $ \rho _{X_{R}}( \varepsilon , \phi )$,
the local density of states at the last layer of the normal, $X_{R}$; (b)
and (d) $J( \varepsilon , \phi )$, the current density. Notice
that $J( \varepsilon , \phi )$, for $| \varepsilon
|<\Delta _{0}$, images the Andreev states: those that contribute to the
current (red and blue---in the online version---are positive and negative
contributions, respectively, given by the sign of the slope $\frac{\partial %
\varepsilon _{j}}{\partial \phi }$). (c) Total Josephson current, $I$, i.e.,
the integral of $J( \varepsilon , \phi )$. Panels (e) and (f):
magnetization, $s_{y}(x,y)\!=\!\frac{\hbar }{2}\langle  \sigma
_{y}\rangle $, at $ \phi_{a}\!=\!0.3 \pi $ for Andreev states
$|A\rangle $ and $|B\rangle $, (see panel (d) and the discussion in Fig.~\ref
{FG:f4}). The magnetizations, integrated over all the normal sample, are
positive: $s_{y}^{B}=0.45s_{y}^{A}$.}\label{FG:f3}
\vspace{-0.4cm}
\end{figure}

Figure \ref{FG:f3} shows the results for zero magnetic field for a device
with strong SO coupling. In Fig.\ref{FG:f3}(a) we show a map of the local
density of states (LDOS) at $x\!=\!x_{R}$ as a function of $\varepsilon $
and $\phi $: $\rho _{X_{R}}(\varepsilon ,\phi )$ (see Appendix \ref{AP:A}).
The non-dispersive states are due to modes that are reflected at the barrier
and do not contribute to the Josephson current density shown in Fig.\ref
{FG:f3}(b). For the gate-voltage shown, the barrier allows the transmission
of a single transverse mode. Thus, at each side of the barrier there are
standing waves that can not transport Cooper pairs from one superconductor to
the other---except for an exponentially small contribution due to quantum
tunneling.

\begin{figure*}[t]
\includegraphics[width=.82\textwidth]{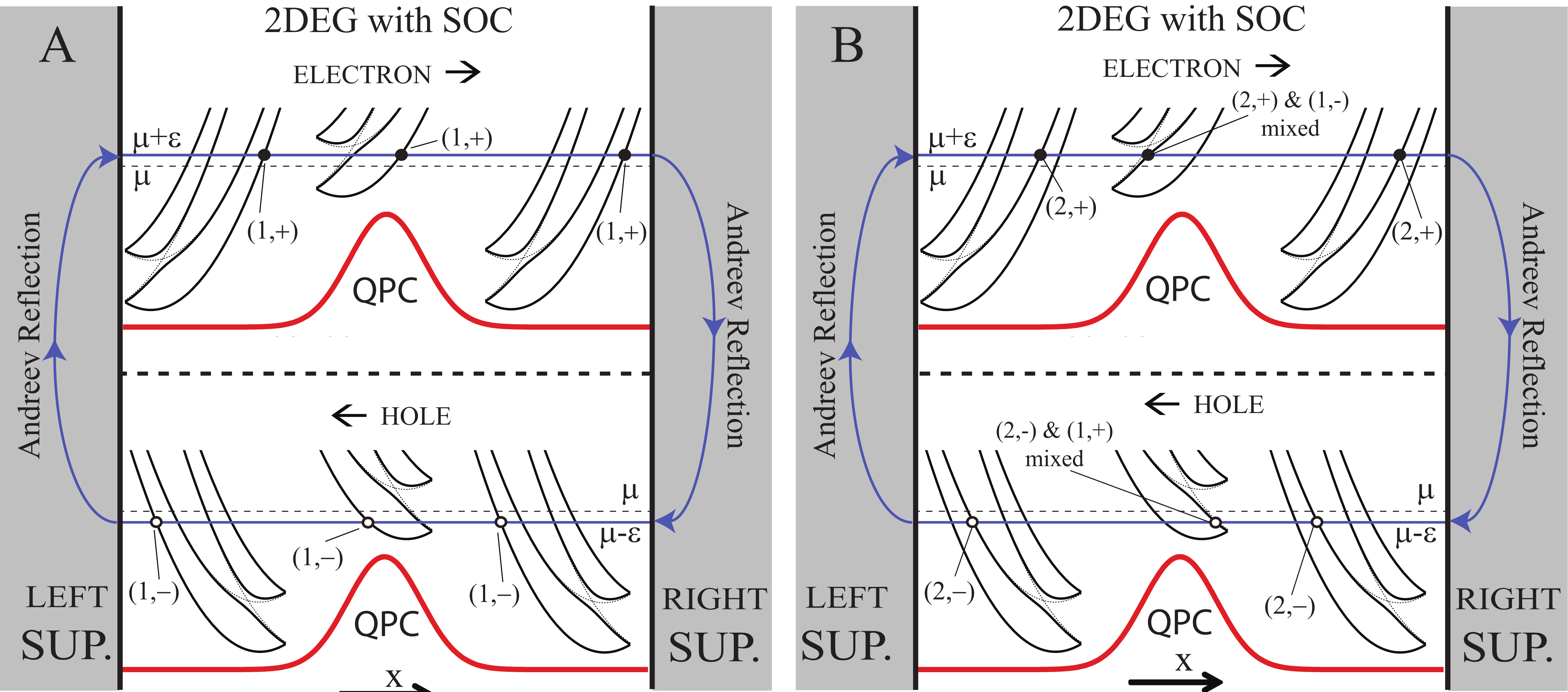}
\caption{(Color online) Sketch for the WKB interpretation of the Andreev
states $|A\rangle $ and $|B\rangle $ shown in Fig.~\ref{FG:f3}(d), (e) and
(f). Both states have positive velocity, they transfer Cooper pairs from the
Left (${L}$) to the Right ($r$) superconductor. Figure \ref{FG:f3} shows
that the state $|A\rangle$ transports Cooper pairs at a faster rate (since $\frac{\partial \varepsilon _{%
\mathrm{A}}}{\partial \phi }>\frac{\partial \varepsilon _{\mathrm{B}}}{ \partial \phi }$) and its $s_{y}$ profile is always positive with a single
maximum along the $y$-direction, whereas $s_{y}$ for the Andreev state $%
|B\rangle $ shows two positive maxima except from the center of the barrier
where there exists a negative contribution to $s_{y}$ having a \emph{single}
maximum. This is in full agreement with the WKB picture presented here. For
the state $|A\rangle $ (left panel) the electron occupies the $(1,+)$
transverse mode all along the barrier as much as the reflected hole occupies
the $(1,-)$ transverse mode. On the other hand for $|B\rangle $ (right
panel), because of the smoothness of the barrier and the avoided crossings
presented in Fig.~\ref{FG:f1}, the electron (hole) is on the $(2,+)$ (($2,-$%
)) mode away from the barrier and it mixes with the $(1,-)$ ($(1,+)$) close
to the barrier. This explains the different patterns of local magnetization
along the $y$-direction, and why the Andreev state $|B\rangle $ is \emph{slower}
than the $|A\rangle $ one.}\label{FG:f4}
\vspace{-0.5cm}
\end{figure*}

On the other hand, the Andreev bound states---those contributing to the
current---have an apparent linear behavior with $\phi$. The sign of their
contributions to $J(\varepsilon ,\phi)$ is correlated with the sign of their
slopes: the latter are linked to the \emph{velocity} of the Andreev states [cf. Eq. %
\eqref{EQ:j}, see Fig.\ref{FG:f3}(b)]. In Fig.\ref{FG:f3}(c) we present the
total Josephson current, $I$, which is given by the integral in Eq.%
\eqref{EQ:iphi0}. It includes the contributions of both the discrete
(Andreev bound states) and of the continuum ($\left| \varepsilon \right|
>\Delta _{0}$) spectrum. The most important contribution comes from the
Andreev bound states. \cite{Andreev64,BlonderTK82} The shape of the CPR for
zero field does not show signatures of the presence of the SOC and the
barrier: similar triangular-like CPR shapes are expected in S-N-S long
junctions with high transparency at the S-N interfaces.\cite
{Kulik69,Bagwell92,Furusaki91,*Furusaki92,Takayanagi95,*Takayanagi95prb,Ebel05}

Even though the magnetic field is zero, the Andreev levels are not
spin-degenerate [Fig.\ref{FG:f3}(d)]. This is due to the Rashba coupling in
the normal region and---as it becomes clearer below---the effect is enhanced
by the presence of the spin polarizer barrier. Let us fix $\phi \!=\!0.3\pi $
and investigate the features of two neighbors Andreev states, $|A\rangle $
and $|B\rangle $. Notice that both Andreev levels have the same slope sign
and therefore they transport Cooper pairs in the same direction. In a system
without magnetism and without SOC, the two states $|A\rangle $ and $%
|B\rangle $ would be degenerate and their spin properties would be opposite
to each other. Here the situation is different: the two Andreev states
neither share the same transversal profile nor their spin-properties are
opposite: in Figs.~\ref{FG:f3}(e) and \ref{FG:f3}(f) we
show their spin density along the $y$-direction---the direction in which the
barrier polarizes the current as a result of the Rashba coupling.

In order to qualitatively analyze the results it is useful to resort to the
WKB picture presented in Sec.\ref{SC:normal}. First, we notice that in the
states $|A\rangle $ and $|B\rangle $ the Cooper pairs are transferred from
the Left (${L}$) superconductor to Right (${R}$) superconductor (positive
slope). Thus, each Andreev bound state is a stationary cyclic sequence of: (%
\textit{i}) electron traveling from ${L}$ to ${R}$, (\textit{ii})
electron-to-hole Andreev reflection at $R$, (\textit{iii}) hole traveling
from ${R}$ to ${L}$, and (\textit{iv}) hole-to-electron Andreev reflection
at ${L}$.

For the subgap Andreev transport ($|\varepsilon|<\Delta_0$), we sketch in
Fig.\ref{FG:f4} the two different possibilities left open by the device as $%
V_g$ is tuned to allow a single transverse mode to transmit. The left panel
shows different processes that lead to the state $|A\rangle$: the electron
travels from ${L}$ to ${R}$ in the $(1,+)$ mode and it is Andreev reflected
as the lack of an electron in the $(1,-)$ channel; then, this hole is
transmitted in the $(1,-)$ channel from ${R}$ to ${L}$ where it is Andreev
reflected as an electron back in channel $(1,+)$ and the loop closes. This
is therefore an state with an overall spin-up polarization in the $y$%
-direction and a transverse spatial profile given by the first transverse
mode (which has a single maximum in the transverse direction). This is in
agreement with the exact results for the spin profile shown in Fig.\ref
{FG:f3}(e).

Similarly, the right panel of Fig.\ref{FG:f4} focuses on the case of the
Andreev state $|B\rangle$; the electron in its travel from ${L}$ to ${R}$
starts at mode $(2,+)$, it mixes with the $(1,-)$ mode in the region of the
barrier and then arrives at $R$ in the $(2,+)$ mode. It is Andreev reflected
as the lack of an electron in $(2,-)$, this hole is transmitted arriving at $%
{L}$ in $(2,-)$ but in the region of the barrier it mixes with mode $(1,+)$.
This is also in agreement with the exact quantum result shown in Fig.\ref
{FG:f3}(f) for state $|B\rangle$, as the local spin density shows two
positive maxima (spin $up$ and second transverse mode) away from the barrier
and a contribution with a single negative maximum (spin $down$ and first
transverse mode) appear in the region of the barrier. Furthermore, within
this picture, it is also justified that the Andreev state $|A\rangle$
transports Cooper pairs at a faster rate than $|B\rangle$ [see greater slope with $\phi$ in Fig.~\ref{FG:f3}%
(d)] because in the first one both electrons and holes are transmitted fully
in the first transverse mode, which is at all positions faster than the
other channels, as it is evident from the slopes $|\frac{dE}{dk}|$ in Fig.%
\ref{FG:f1}. \vspace{-0.3cm}

\subsubsection{Symmetry considerations and the weak magnetic field limit}
\label{SC:SymCPR}
We start studying the symmetry properties of the BdG Hamiltonian, $H_{BdG}$,
in absence of ferromagnetic materials and magnetic fields. In this case, we
have
\begin{equation}
H_{BdG}(-\phi )=\mathcal{T}^{-1}H_{BdG}(\phi )\mathcal{T}\,,
\label{EQ:simTRS}
\end{equation}
where $\mathcal{T}$ is the time-reversal operator Notice that $-\phi $ is
equivalent to $2\pi -\phi $. With the exceptions of $\phi \!=\!0$ and $\phi
\!=\!\pi $, Eq.\eqref{EQ:simTRS} does not imply the existence of degenerate
solutions for $H_{BdG}$ at a fixed $\phi $. Instead, it allows us to
construct solutions for the case of $\phi \!=\!(2\pi -\phi _{x})$ from the
ones at $\phi \!=\!\phi _{x}$. Thus, for any given eigenstate (labeled by
the natural number ${j}\!=\!1,2,...$) $|j\rangle $ at $(\varepsilon ,\phi
)=(\varepsilon _{j},\phi _{x})$, we know that
\begin{equation}
\left.
\begin{array}{c}
|{j}\rangle  \\
(\varepsilon _{j},\phi _{x})
\end{array}
\right. \Rightarrow \left.
\begin{array}{c}
|{{\bar{j}}}\rangle \!\equiv \!\mathcal{T}^{-1}|{j}\rangle  \\
(\varepsilon _{j},2\pi -\phi _{x})
\end{array}
\right. .  \label{EQ:trsll}
\end{equation}
This means that if the energy of $|j\rangle $ at $\phi \!=\!\delta \phi
+\phi _{x}$ is given by $\varepsilon _{j}(\delta \phi )\!=\!\varepsilon
_{j}(\phi _{x})+b_{j}\,\delta \phi $ then the energy of $|{{\bar{j}}}\rangle
$ at $\phi \!=\!\delta \phi +(2\pi -\phi _{x})$ is $\varepsilon _{{{\bar{j}}}%
}(\delta \phi )\!=\!\varepsilon _{j}(\phi _{x})\!-\!b_{j}\,\delta \phi $,
with $b_{j}$ the slope of the Andreev level for $|j\rangle $ at $\phi =\phi
_{x}$. Therefore, their velocities, $v_{j}$ and $v_{{{\bar{j}}}}$, become
opposite to each other ($v_{j}\!=\!-v_{{{\bar{j}}}}$) as they are
proportional to $\frac{\partial \varepsilon _{j}}{\partial \phi }$ and $%
\frac{\partial \varepsilon _{{{\bar{j}}}}}{\partial \phi }$, respectively.
This implies that the Josephson current must be zero for either $\phi
_{x}\!=\!0$ or $\phi =\pi $, as the two $\mathcal{T}$-related states share
the same coordinates $(\varepsilon ,\phi )$ so their individual
contributions cancel out. For arbitrary values of $\phi $ the following
relations hold
\bese
\label{EQ:Isymm}
\begin{eqnarray}
I(\phi ) &=&-I(-\phi )=-I(2\pi -\phi ), \\
I(\pi +\phi ) &=&-I(\pi -\phi ).  \label{EQ:Ipi}
\end{eqnarray}
\eese
An obvious property implied by Eq.\eqref{EQ:Isymm} is that the absolute
value of the critical current is independent on the direction of the
current, i.e. $I_{c}^{+}=I_{c}^{-}$ with
\begin{equation}
I_{c}^{+}\equiv \max I(\phi )~,~~I_{c}^{-}\equiv \left| \min I(\phi )\right|
.
\end{equation}
Over the next Sections, we discuss how the barrier presented in Sec.\ref
{SC:normal} provides a controllable way to violate the relation $%
I_{c}^{+}\!=\!I_{c}^{-}$. In this section we discuss the fate of the
properties Eq.\eqref{EQ:Isymm} when Eq.\eqref{EQ:simTRS} no longer holds due
to the presence of external magnetic fields coexisting with spin-orbit
coupling.

Starting from the zero magnetic field condition we group the Andreev states
in $\mathcal{T}$-related states, $|{j}\rangle$ and $|{{\bar{j}}}\rangle$,
and address the response of the energy levels to a weak external magnetic
field. As $|{{\bar{j}}}\rangle$ is the time-reversed state of $|{j}\rangle$,
they have opposite local spin projections,
\begin{equation}
\langle \bm{s}_{{j}}({\bm r})\rangle\equiv\langle {j}|\bm{s}({\bm r})|{j}%
\rangle=-\langle {{\bar{j}}}|\bm{s}({\bm r})|{{\bar{j}}}\rangle=-\langle %
\bm{s}_{{{\bar{j}}}}({\bm r})\rangle\,,  \label{EQ:spinTQpairs}
\end{equation}
with $\bm{s}\! = \! \frac{\hbar}{2}{\bm{\sigma}}$ and $\bm{\sigma}\! = \!(
\sigma_x,\sigma_y,\sigma_z)$. Let us assume, as an example, that the spin of
the Andreev state $|{j}\rangle$ is \emph{up} along the $y$-direction---then
we know that $|{{\bar{j}}}\rangle$ is a \emph{down} state. Thus, if a Zeeman
field is applied along the $y$-direction---provided it is not too strong to
affect the properties of the original states---the energy levels
corresponding to $|{j}\rangle$ (at $\phi\! = \!\phi_x$) and to $|{{\bar{j}}}%
\rangle$ (at $\phi\! = \! 2\pi-\phi_x$) change in the same amount, $%
|\varepsilon_Z|$, but with opposite signs. In general, if the total spin of
the Andreev state is not fully aligned with the external magnetic field then
the splitting will be a fraction of $|\varepsilon_Z|$ (but still
proportional to the magnetic field).
We must also keep in mind that the states change their spin structure as $%
\phi$ (and $\varepsilon$) varies.

\begin{figure}[t]
\includegraphics[width=.48\textwidth]{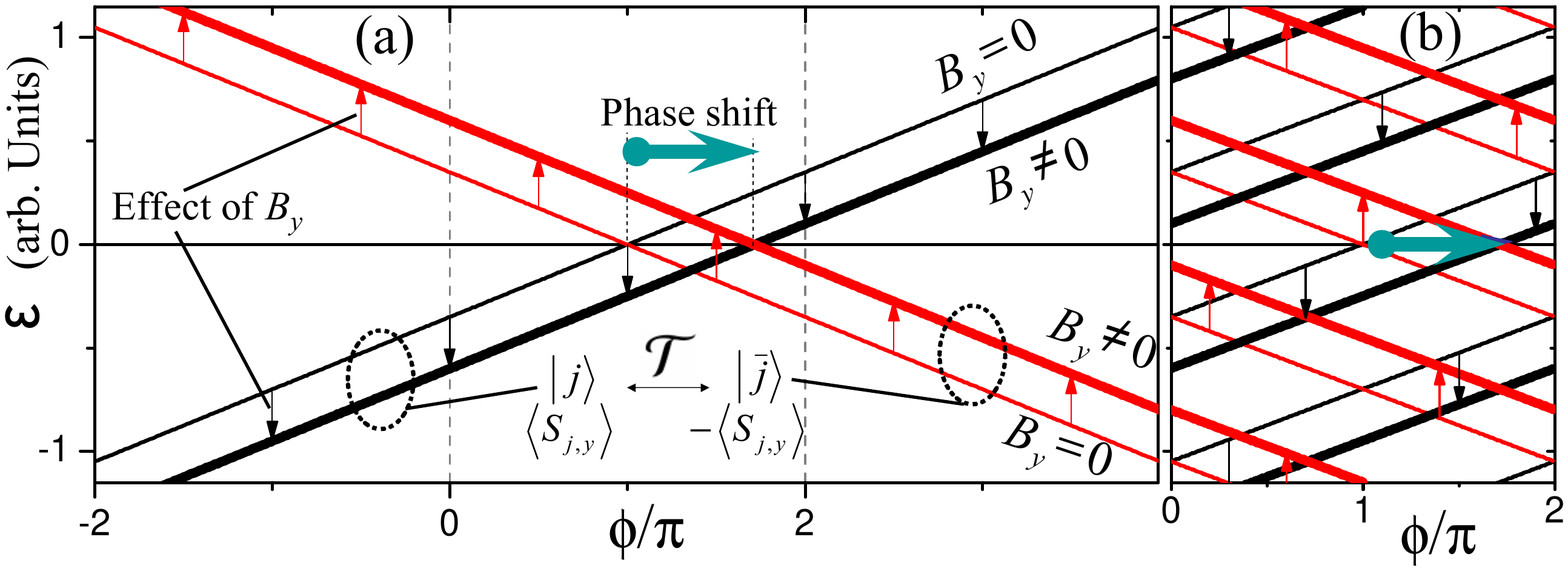} \vspace{-0.6cm} 
\caption{(Color online) Phase shifts due to external magnetic fields. The
N-S interfaces are transparent. (a) Andreev levels in the long junction
limit. At zero field, due to the symmetry under $\mathcal{T}$ of the BdG
Hamiltonian, for every state $|{j}\rangle$, at $ \phi_x$, there is an
associated state $|{{\bar{j}}}\rangle$, at $2 \pi- \phi_x$ ($%
\! \equiv \!- \phi_x$) that: (i) has the same energy, (ii) has
opposite local spins properties (see text), and (iii) has opposite velocity.
A weak in-plane magnetic field $B_y$---if applied along a direction in which
the total spin of the Andreev state is nonzero---generates an increment
(decrement) of the energy of the state $|{{\bar{j}}}\rangle$ ($|{j}\rangle$) as
shown with vertical arrows. The new pattern of Andreev levels (thick lines)
are phase-shifted (horizontal arrow) from the $B\! = \!0$ levels (thin
lines). (b) Shows the Andreev levels of (a) restricting $ \phi$ to
the $(0,2 \pi)$ region.}
\label{FG:f5}
\end{figure}

In order to qualitatively understand how phase-shifts may arise due to the
Zeeman energy shift, $\varepsilon_Z$, we now particularize to the case of a
transparent N-S interfaces in a long junction---a situation in which their
Andreev levels can be approximated as linear functions of the phase for $%
|\varepsilon|<\Delta_0$.\cite{Kulik69,Bagwell92} For $|\varepsilon|<\Delta_0$
we write the levels as
\bese
\label{EQ:shiftE}
\begin{eqnarray}
\varepsilon_{j}(\phi,\varepsilon_Z)&=&\lambda_{j} (\phi-\pi) - \varepsilon_Z
\frac{2}{\hbar}\langle S_{{j},y}\rangle\,, \\
\varepsilon_{{{\bar{j}}}}(\phi,\varepsilon_Z)&=&-\lambda_{j} (\phi-\pi) -
\varepsilon_Z \frac{2}{\hbar} \langle S_{{{\bar{j}}},y}\rangle\,, \\
&=& -\lambda_{j} (\phi-\pi) + \varepsilon_Z \frac{2}{\hbar} \langle S_{{j}%
,y}\rangle\,,
\end{eqnarray}
\eese
where $\lambda_{j}>0$ is the slope of the level and $\langle S_{q,y}\rangle$
is the total spin along the $y$-direction of the Andreev state $|q\rangle$%
---we later relate these quantities with the parameters of our system. From
Eq.\eqref{EQ:shiftE} we see that the shift in energy, due to the presence of
the magnetic field $B_y$, is equivalent to a shift of the superconducting
phase of the Andreev levels at zero field,
\bese
\label{EQ:shiftPhi}
\begin{eqnarray}
\varepsilon_{{j}}(\phi,\varepsilon_Z)&=&\varepsilon_{j}(\phi-\phi_{j}^Z,0)\,,
\\
\varepsilon_{{{\bar{j}}}}(\phi,\varepsilon_Z)&=&\varepsilon_{{{\bar{j}}}%
}(\phi-\phi_{j}^Z,0)\,, \\
\phi_{{j}}^Z&=& \frac{2 \varepsilon_Z}{\hbar \lambda_{j}}\langle S_{{j}%
,y}\rangle \,.
\end{eqnarray}
\eese
The energy levels in Eq.\eqref{EQ:shiftE} and Eq.\eqref{EQ:shiftPhi} are
single-valued functions in $\phi$. However, by translating all solutions to
the $[0,2\pi]$ region, we get many Andreev levels for each value of $\phi$
in that range. The effect of the magnetic field is shown in Fig.\ref{FG:f5}%
(a) for an unrestricted phase axis while Fig.\ref{FG:f5}(b) shows the same
results folded into the $[0,2\pi]$ region. This illustrates how the energy
levels of a $\mathcal{T}$-pair of Andreev states would rearrange due to the
magnetic field in any long junction in the transparent interface limit.

From the above discussion we see that the $\mathcal{T}-$pairs provide a
useful way to group the Andreev levels when discussing how a weak magnetic
field can modify them. It is appropriate to make an important remark here. It
is known that for a system with zero spin-orbit coupling but nonzero
magnetic field the Josephson current fulfills the constrictions of Eq.%
\eqref{EQ:Isymm} irrespective of the transparency of the interfaces, the
value of the magnetic field and the length of the junction. This follows
from a symmetry of the Hamiltonian analog to the one presented in Eq.%
\eqref{EQ:simTRS} for time-reversal systems, namely,\cite{Margaris2012}
\begin{equation}
H_{BdG}(-\phi)= \mathcal{U}^{-1} H_{BdG}(\phi)\mathcal{U}\,,  \label{EQ:sim2}
\end{equation}
where $\mathcal{U}=\mathcal{T}R_{\pi}$ and $R_{\pi}$ is a $\pi$-rotation of
the spin along an axis perpendicular to the one of the magnetic field. The
TR operator, $\mathcal{T}$, reverses the magnetic field direction and $%
R_{\pi}$ maps the field back to its original direction. The relation holds
even with a position dependent magnetic field, providing the magnetic field
is coplanar---the rotation $R_{\pi}$ is taken along the normal to such
plane. The properties of Eq.\eqref{EQ:Isymm} for the CPR follow
straightforwardly by noticing that the two states in a $\mathcal{U}-$pair
(analogous to states in the $\mathcal{T}-$pair of Eq.\eqref{EQ:trsll}) have
opposite velocities.

The important point is that, for Josephson junctions in which SOC and Zeeman
fields coexists, a symmetry operator $\mathcal{U^{\prime}}$ as the one in Eq.%
\eqref{EQ:sim2} does not exists in general and thus the CPR properties shown
in Eq.\eqref{EQ:Isymm} can be violated: the CPR can have $I(0)\! \neq \! 0$,
$I(\pi)\! \neq \! 0$ and $I^{+}_c \! \neq \! I^{-}_c$.\cite{Liu2010} It is
instructive to discuss, for example, the case of 1D spin-orbit system with a
Zeeman field,
\begin{equation}
\mathcal{H}^0\! = \!\frac{1}{2m^*}(p_x -\hbar k_\alpha\sigma_1)^2 - E_\alpha
+E_{Z,1}\sigma_1 + E_{Z,3}\sigma_3 \,,  \label{EQ:H1dsocZeeman}
\end{equation}
where we have written the $\{1,2,3\}$-directions in spin space for gaining
generality ($\sigma_i$ are the Pauli matrices) with the SOC along the $1$%
-direction. $E_{Z,1}$ and $E_{Z,3}$ are the parallel and perpendicular
Zeeman energies, respectively. This Hamiltonian can be mapped by a
well-known unitary transformation to a system without SOC but with an
inhomogeneous magnetic field. The transformation is just the spin-dependent
shift in momentum,\cite{AleinerFalko2001} $U_\alpha\! = \!\exp\left(\mathrm{i%
} k_\alpha x \sigma_1\right)$. In the transformed Hamiltonian, the parallel
magnetic field is unaffected, whereas the perpendicular component rotates
along the $1-$axis as a function of the position $x$,
\begin{eqnarray}
U_\alpha^\dagger\mathcal{H}^0 U_\alpha^{} &= &\frac{p_x^2}{2m^*} - E_\alpha
+E_{Z,1}\sigma_1  \notag \\
&& + E_{Z,3}\left[ {\cos \left(2k_\alpha x\right) \sigma_3 - \sin \left(2
k_\alpha x\right) \sigma_2}\right]\,.
\end{eqnarray}
Notice also that the s-wave superconducting pairing potential transforms
trivially under $U_\alpha$.\cite{MortenKarsten2011} This simple argument
allows us to see that for magnetic fields that are neither parallel nor
perpendicular to the SOC axis---i.e., $E_{Z,1}\! \neq \! 0$ and $E_{Z,3}\!
\neq \! 0$---a symmetry as in Eq.\eqref{EQ:sim2} cannot be found because the
system is intrinsically equivalent to a system without SOC but subject to a
non coplanar Zeeman field texture.

For the 2D case with Rashba SOC, it has been shown in Ref.~\onlinecite{Liu2010} that a symmetry $\mathcal{U^{\prime}}$ that assures
zero Josephson current for $\phi\! = \!0$ can be found only if the Zeeman
interaction is perpendicular to the plane of the SOC.\footnote{%
Except for the special case of identical linear-Dresselhaus and Rashba
strengths for which some in-plane directions can also lead to a
symmetry-blockade of the current for $\phi\! = \!0$.} For inplane fields, as
Eq.\eqref{EQ:Isymm} no longer holds, current for $\phi\! = \!0$ and $\phi\!
= \! \pi$ is not forbidden by symmetry and also one may find that $I^{+}_c\!
\neq \! I^{-}_c$. However, the geometry of the sample plays a crucial role
since not any in-plane direction is efficient to modify the Josephson
current. For instance, due to the sample configuration (width going to
infinity) phase-shifts in Ref.~\onlinecite{Bezuglyi02} are not observed
for both directions of in-plane magnetic fields. Similarly, in the finite
width stripe geometry we investigate,\cite{ReynosoUB08prl} (with or without
a symmetric barrier inside) if the magnetic field is \emph{parallel} to the
stripe the current for $\phi\! = \!0$ becomes zero.

From the above discussion we have deepened our physical understanding of the
presence of a finite Josephson current at zero phase difference. Our
TR-symmetric device is able to break the spin degeneracy of the Andreev
states in absence of magnetic fields. Due to the sample-specific geometry, a
given in-plane direction is privileged in the spin of the Andreev states.
More specifically, with a single transverse mode open inside the barrier the
two Andreev states with positive velocity ($|A\rangle$ and $|B\rangle$
states in Figs.\ref{FG:f3} and \ref{FG:f4}) can have the \emph{same} sign of
total magnetization \emph{along} the direction in which the SOC device
polarizes. From the concepts that led us to Eq.\eqref{EQ:shiftPhi} it follows that a
nonzero current for zero phase difference is expected if a weak magnetic
field $B_y$ is applied as $\phi_{A}^Z\! \neq \! -\phi_{B}^Z$. On the other
hand, current for zero phase difference is not expected for inplane magnetic
fields along the $x-$direction because for those states the integrated
magnetization along $x$-direction is zero, i.e., $\langle S_{A,x}\rangle\! =
\! \langle S_{B,x}\rangle \! = \! 0$. This is in agreement with our
numerical simulations.

\subsubsection{Andreev levels and phase-shifts in the weak magnetic field
limit}

\label{SC:phaseShift}

In order to analyze our numerical results quantitatively, we now relate the
two essential features of the Andreev states in transparent junctions,
namely their phase dependence (slope) and their energies as a function of a
magnetic field (in linear response), with the parameters of our system. The
energies of Andreev states for a homogeneous and long (${L_N}\gg \xi
_{N}\equiv \hbar v_\mathrm{F}/\pi \Delta $) one-dimensional S-N-S junction
are given by $\varepsilon ^{n,\pm }=\hbar v_\mathrm{F}/(2{L_N})[2\pi
(n+1/2)\pm \phi ]$ where $v_\mathrm{F}$ is the Fermi velocity of the normal
material, ${L_N}$ is the length of the junction and the plus and minus signs
correspond to the two signs of the excitation velocity. This expression can
be generalized to include both the effect of external field and corrections
of order $\xi _{N}/{L_N}$. Following the procedure of Ref.~\onlinecite{Kulik69} we obtain
\begin{equation}
\varepsilon _{\sigma }^{n,\pm }=\sigma \mu _{\mathrm{eff}}B_{y}+\frac{\hbar
v_\mathrm{F}}{2({L_N}+\pi \xi _{N})}[2\pi (n+1/2)\pm \phi]\,,
\label{EQ:level0}
\end{equation}
here $\sigma $ is the spin of the Andreev state and $\mu _{\mathrm{eff}}$ is the effective magnetic
moment
\begin{equation}
\mu _{\mathrm{eff}}=\frac{g_{N}\mu _{B}}{2}\left( 1+\frac{\frac{g_{S}}{g_{N}}%
-1}{1+\frac{{L_N}}{\pi \xi _{N}}}\right)\,.  \label{ECjj:mueff}
\end{equation}
We assume that the magnetic field is in the $y-$%
direction---since it is contained in the plane of the 2DEG no diamagnetic
effects are induced. The energies of the Andreev states are then simply
shifted by the field. In Eq.\eqref{EQ:level0} $n$ is an integer and $\phi$ is
restricted to the $[0,2\pi]$ region. These energy levels are equivalent to
the ones presented in Eq.\eqref{EQ:shiftE} after folding them into the $[0,2\pi]$ interval, as done in
Fig.~\ref{FG:f5}(b).

This analytical result for a 1D system without SOC can be used to
qualitatively analyse the effects induced by the presence of a barrier in a normal region
when $\alpha\! \neq \!0$. We resort to the WKB approximation and proceed in
the following way: for an adiabatic QPC in the absence of external magnetic
fields and for an electron (and the reflected hole) in the channel $i$ we define
a `local' wavevector,
\begin{equation}
k_{i}^{e/h}(x)=F_{i}^{e/h}(x,\varepsilon,\mu)\,.
\end{equation}
These functions have to be obtained from the dispersion relations at each
position $x_a$, by assuming that the potential for all $x$ is equal
to $V(x_a,y)$. As shown in Sec.\ref{SC:normal}, due to the SOC the different
transverse modes with dispersion relations given by Eq.\eqref{EQ:disp}
become mixed and the form of $F_{i}^{e/h}(x,\varepsilon,\mu)$ become
nontrivial---in the absence of SOC, different transverse modes do not mix
and the functions are just $F_{i}^{e/h}(x,\varepsilon,\mu)\! = \!\sqrt{\frac{%
2m^{\ast }}{\hbar ^{2}}(\mu \pm \varepsilon )-\widetilde{V}_{i}(x)}$, with $%
\widetilde{V}_{i}(x)$ the effective 1D potential that accounts for the shift
of the bottom of the $i$-channel band.

The orbital phase accumulated by the electron and the reflected hole in the
junction is given by
\begin{equation}
\int_{0}^{{L_N}}(k_{i}^{e}(x)-k_{i}^{h}(x))\,dx=\frac{2{L_N}\delta
_{i}\varepsilon }{\hbar v_\mathrm{F}}\,,
\end{equation}
where we have parametrized the integral by a single dimensionless number $%
\delta _{i}$. With this parametrization we find for the energy of the
Andreev states
\begin{equation}
\varepsilon _{i,\sigma }^{n,\pm }=\frac{\hbar v_\mathrm{F}}{2({L_N}\delta
_{i}+\pi \xi _{N})}[2\pi (n+1/2)\pm \phi]\,.
\end{equation}
Within this picture, the effect of the barrier only introduces an effective
length ${L_N}\delta _{i}>{L_N}$ for each channel---the value of $v_\mathrm{F}
$ is taken far away from the barrier potential. There is also a small correction to the effective magnetic moment of channel
$i$ that is given by Eq.~\eqref%
{ECjj:mueff} with ${L_N}$ replaced by the effective length ${L_N}\delta
_{i} $.

\begin{figure}[b]
\includegraphics[width=.48\textwidth]{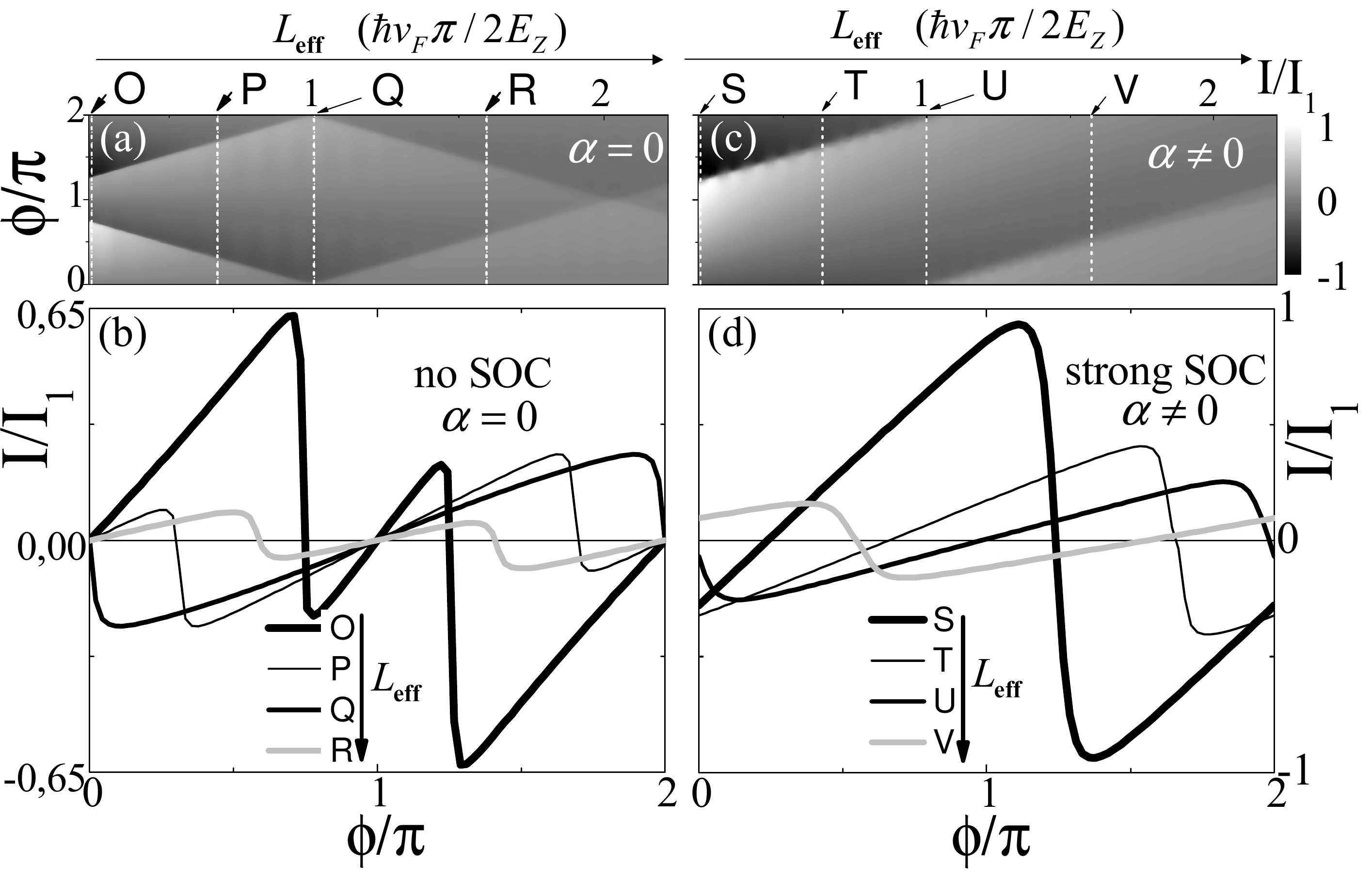} \vspace{-0.7cm}
\caption{(Color online) Josephson current at a fixed finite Zeeman energy, $%
E_Z\! = \! \mu_\mathrm{eff} B_y$, and a fixed barrier profile---$V_g$
is set to allow transmission of a single transverse mode---as ${L_N}$ is
enlarged. For convenience we plot it as a function of the effective length, $L_%
\mathrm{eff}\! = \! \delta_j{L_N}$ (see Eq.\eqref{EQ:shiftA} for $%
 \xi_N\ll {L_N}$). Panels (a) and (b) correspond to $ \alpha\! = \!
0 $, whereas panels (c) and (d) to $ \alpha\! \neq \!0$.
Qualitatively similar results are obtained if ${L_N}$ is fixed and $E_Z$ is
varied. For $ \alpha\! = \! 0$ the normal device behaves as a
ferromagnetic polarizer; the current at $ \phi\! = \!0$ is zero,
while changing the length produce $0$-$\pi$ junction transitions.
On the other hand, for the case shown for $ \alpha\! \neq \! 0$, the
QPC behaves as an efficient polarizer in the normal state. The magnetic field $B_y$ is applied along
the SOC polarization axis. Note that the current at $ \phi\! = \!0$ is nonzero,
and that the phase-shift changes with the length of the junction.}
\label{FG:f6}
\end{figure}

We found, by inspecting the Andreev levels's slopes of the exact numeric
results, that the parameters $\delta_{i}$ are only weakly dependent on the
SOC (see below). As the spin is no longer a good quantum number, in the
above expression the spin index should be replaced by a new quantum number
that characterizes the two branches of each of the $i=1,2,...,N_{oc}$
transmitted channels. In order to include the effect of an external magnetic
field in our simple WKB picture, we profit from the properties of the
Andreev states introduced in the previous section. We group the Andreev
states in $\mathcal{T}$-partners, $|{j}\rangle$ and $|{{\bar{j}}}\rangle$,
given in Eq.\eqref{EQ:trsll} [see Fig.~\ref{FG:f5}], where $({j},{{\bar{j}}}%
)\! = \!(1,\bar{1}),(2,\bar{2}),...,(2 N_{oc}, \overline{2 N_{oc}})$, and ${j%
}$ (${{{\bar{j}}}}$) labels the states with positive (negative) slope.
Therefore, in the linear response regime, a magnetic field $B_{y}$ leads to
the phase shifted energy levels of Eq.\eqref{EQ:shiftE}. In fact, the WKB
version of Eq.\eqref{EQ:shiftE} can be written as
\bese \label{EQ:shiftPhi2}
\begin{eqnarray}
\varepsilon _{j}(\phi ) &=&\eta _{j}\mu _{\mathrm{eff}}B_{y}+\frac{\hbar v_%
\mathrm{F}}{2({L_N}\delta _{j}+\pi \xi _{N})}[2\pi (n +1/2)+\phi ]\,,
\label{andeev 2} \\
\varepsilon _{{{\bar{j}}}}(\phi ) &=&-\eta _{{j}}\mu _{\mathrm{eff}}B_{y}+%
\frac{\hbar v_\mathrm{F}}{2({L_N}\delta_{{j}}+\pi \xi _{N})}[2\pi (n
+3/2)-\phi ]\,,~~~~~~~
\end{eqnarray}
\eese
where the parameter $|\eta_{j}|
\leq 1$ quantifies how different from the spin degenerate case the system
behaves. The link between the parameters and the phase-shift $\phi_{{j}}^Z$
defined in Eq.\eqref{EQ:shiftE} can be written as
\bese
\label{EQ:levels}
\begin{eqnarray}  \label{EQ:shiftPhi3}
\lambda_{j} &=& \frac{\hbar v_\mathrm{F}}{2({L_N}\delta _{j}+\pi \xi _{N})}%
\,,  \label{EQ:slope} \\
\left|\eta _{j}\mu _{\mathrm{eff}}B_{y}\right| &=& \left|\varepsilon_Z \frac{%
2}{\hbar} \langle S_{{j},y}\rangle\right|\,, \\
\phi_{j}^Z&=& {\eta _{j}\mu _{\mathrm{eff}}B_{y}} \frac {2({L_N}\delta _{{j}%
}+\pi \xi_{N})}{\hbar v_\mathrm{F}}\,.  \label{EQ:shiftA}
\end{eqnarray}
\eese
The factor $\pi \xi _{N}$ in the denominator can
be neglected in the long junction limit---this is the case of the results presented in Fig.~\ref{FG:f6}
that we discuss below. Introducing the linear dispersions of Andreev levels
in Eqs.\eqref{EQ:iphi0} and \eqref{EQ:j}, the zero magnetic field and zero
temperature contribution to the Josephson current of each 1D mode (a set of $%
{j}$ and ${{\bar{j}}}$ partners) is proportional to the saw-tooth function, $%
F_{\mathrm{st}}(\phi)$,\cite{Kulik69,Bagwell92}
\begin{equation}
I_{j}(\phi)=\frac{2e \lambda_{j}}{\hbar} F_{\mathrm{st}}(\phi)=\left\{
\begin{array}{l}
\frac{2e \lambda_{j}}{\hbar} \frac{\phi}{\pi}~~~~~~~~~~~~~\mathrm{if~}%
0<\phi\leq \pi \\
\frac{2e \lambda_{j}}{\hbar}{\left(\frac{\phi}{\pi}-2\right)}~~~\mathrm{if~}%
\pi<\phi\leq 2\pi
\end{array}
\right. \,,  \label{EQ:II}
\end{equation}
which is a $2\pi$-periodic zero-mean function and we have defined it for $%
\phi\in[0,2\pi]$. For $\frac{{L_N}}{%
\xi_{N}}$ finite the total current includes the contributions of the Andreev bound states ($|\varepsilon|<\Delta_0$) and of the continuum states ($|\varepsilon|>\Delta_0$).\cite{Bagwell92} For a perfect 1D channel the CPR has an abrupt change at $\phi\! = \!\pi$, however, in the simulation such jump is slightly smoothed due to the presence of the barrier (see Fig.\ref{FG:f3}(c)).

\begin{figure*}[!t]
\includegraphics[width=.97\textwidth]{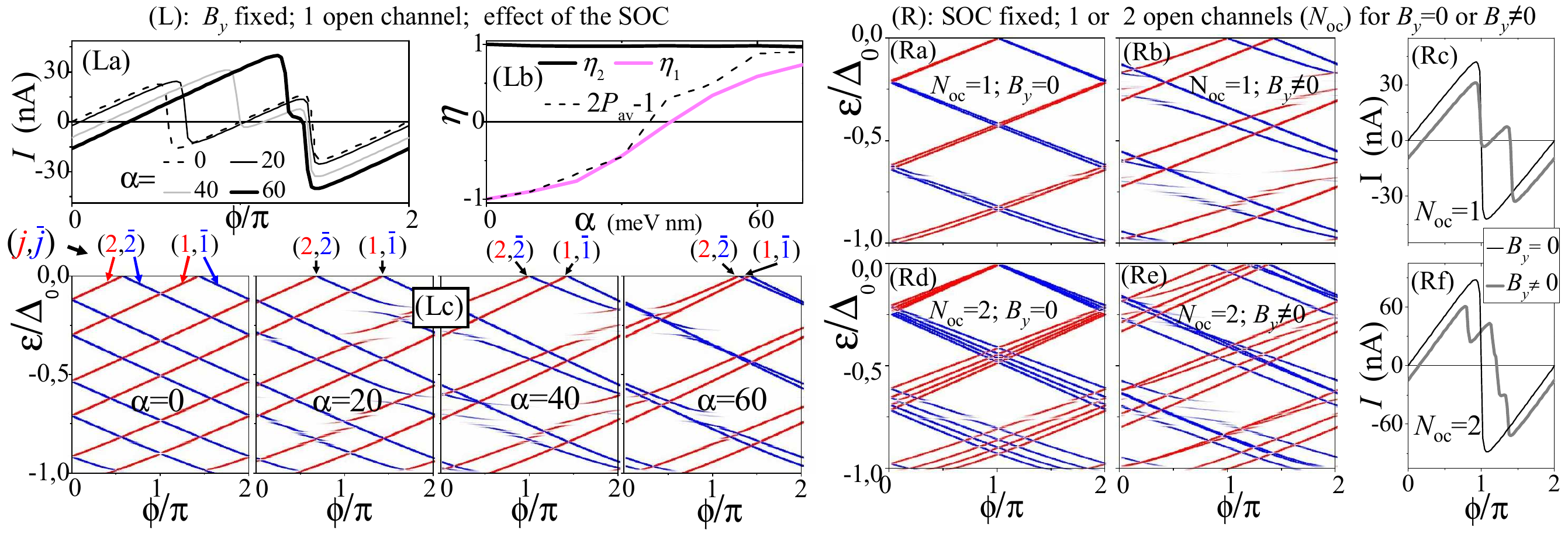} \vspace{-0.4cm}
\caption{(Color online) Anomalous current, $I( \phi\! = \! 0)\! \neq
\! 0$, phase-asymmetric CPRs and current-asymmetric CPRs, due to in-plane
magnetic field (parallel to the SOC polarization axis). The barrier is the
quantum point contact, $\mathrm{QPC}_1$, described by Eq.\eqref{EQ:potential}
taking $(z;W_b; L_b)\! = \!(30; 100; 250)$nm. The junction length is ${L_N}%
\! = \!$ 1.2$ \mu$m and the S-N interfaces are transparent. The Left
(L) panels show CPRs (La) and Andreev levels (Lc) for a fixed magnetic
field, $E_Z\! = \! g  \mu_B B_y \! = \!\frac{\Delta_0}{10}\! = \! 150%
 \mu$eV, and different values of the Rashba strength, $ \alpha$%
. The gate voltage, $V_g\! = \! 46.5$meV, sets the number of open transverse
channels, $N_{oc}$, in $1$. The larger $ \alpha$ the bigger is the
current $I( \phi\! = \!0)$ and the less distorted is the Josephson
current from the one at $B_y\! = \!0$: it looks as a phase-shifted version
of the CPR at zero field (see (Ra) and (Rc)). (Lc) Deduced relative magnetic
moments, $ \mu_1$ and $ \mu_2$, of the two Andreev bound
states states (see text) as a function of $ \alpha$. We show the
simplest estimation for $ \mu_2$, $(2 P_{av}-1)$, with $P_{av}$ the
average polarization induced by the device in the energy range of the
superconducting gap. The Right (R) panels show how the opening of more than one
transverse mode in the QPC can lead to asymmetric CPRs when $B_y$ is
nonzero. We fix $ \alpha\! = \! 40$meVnm and show the Andreev levels,
for $(V_g,N_{oc})\! = \! (46.5\mathrm{meV},1)$ [~$(39.4\mathrm{meV},2)$] in
panel (Ra) [(Rd)] for $B_y\! = \!0$ and in panel (Rb) [(Re)] for $B_y\! \neq
\!0$: $E_Z\! = \! 150 \mu$eV. In panel (Rf) the CPR for $B_y\! \neq
\!0$ and $N_{oc}\! = \!2$ does show current-asymmetry, i.e., it fulfills Eq.
\eqref{EQ:Isymm}. On the other hand for the case of $N_{oc}\! = \! 1$ shown
in panel (Rc) the CPR gets phase-shifted and distorted for $B_y\! \neq \!0$
but no strong current-asymmetry is observed.}
\label{FG:f7}
\end{figure*}

We now discuss the effect of the magnetic field on the Josephson current. The
gate voltage $V_g$ is set to allow the transmission of a single channel,
i.e., $N_{oc}\! = \!1$. Thus, there are two distinct Andreev levels, $({j},{{\bar{j}}})=(1,\bar{1}),(2,\bar{2})$,
with positive and negative slope, respectively. In the
simulation there is a weak in-plane magnetic field (i.e., $E_Z\ll \Delta_0$)
and we vary the length of the junction while the barrier dimensions
are kept fixed. The quantities $\delta_{{j}} {L_N}$ change and
so the phase shifts in Eq.\eqref{EQ:shiftPhi3}. The current becomes
\begin{equation}
I(\phi)=\frac{2e \lambda_{1}}{\hbar} F_{\mathrm{st}}(\phi-\phi_{1}^Z)+\frac{%
2e \lambda_{2}}{\hbar} F_{\mathrm{st}}(\phi-\phi_{2}^Z)\,.  \label{EQ:Ist}
\end{equation}
We show the CPRs for a system without SOC in Fig. \ref{FG:f6}(a) and \ref
{FG:f6}(b). Note that there is no current for $\phi\! = \!0$ because the
states ${j}=1$ and ${j}=2$ are the spin \emph{up} and \emph{down} states
along the $y$-direction, respectively, and so $\phi_{2}^Z\! = \! -\phi_{1}^Z$
(because $\eta_{2}\! = \! -\eta_{1}\! = \! 1$). For $\phi_{1}^Z\! =
\!-\phi_{2}^Z\! = \! (2n+1)\pi$ the discontinuity of both saw-tooth
functions is at $\phi\! = \!0$. This can be understood as a transition from
a $0$ to a $\pi-$junction.

Figure \ref{FG:f6}(c) and \ref{FG:f6}(d) show the CPR for a system with
polarizing properties due to the presence of strong SOC. The current for $%
\phi\! = \!0$ is nonzero as both saw-tooth components move in the same
direction. This can be understood as follows: as the mode $(1,+)$ dominates the Andreev state  $%
|A\rangle $ [see Fig.~\ref{FG:f3}(e) and Fig.~\ref{FG:f4}(A)] we have $%
\eta_{1}\! = \!\eta_{\mathrm{A}} \! \approx \! 1$. On the other hand, since the region
of the barrier is small compared to the total length of the junction, the $(2,+)$ mode dominates away from the barrier in the Andreev state $%
|B\rangle$ [see Fig.~\ref{FG:f3}(f) and Fig.~\ref{FG:f4}(B)], and we also have $%
\eta_{2}\! = \!\eta_{\mathrm{B}} \! \approx \! 1$. It then follows that $%
\phi_{2}^Z\! \approx \! \phi_{1}^Z$ justifying the numerical results of Fig.
\ref{FG:f6}(c). Using Eqs.\eqref{EQ:Ist}, \eqref{EQ:II} and %
\eqref{EQ:shiftPhi2} we see that the longer the junction the bigger the
phase-shift ($\propto {L_N}$) while the maximum current gets reduced
(because $\lambda_{j} \propto {L_N}^{-1}$). Similarly, if ${L_N}$ is fixed
both the phase-shift and $I(\phi\! = \! 0)$ grow linearly with $B_y$ while
the critical current does not change.

\subsubsection{Phase shift of Andreev levels, dependence with $ %
\alpha $}

In Figure \ref{FG:f7}(L) we show the Andreev levels and the CPRs for several values of $\alpha$
corresponding to the weak, intermediate and the strong SOC regime. The number of open channels, $%
N_{oc}$, is one and both the junction length and the magnetic field are
fixed. As expected, for $\alpha\! = \!0$ [Fig.~\ref{FG:f7}(Lc)] the
arrangement of Andreev levels is symmetric with respect to $\phi\! = \!\pi$.
We call $({j},{{\bar{j}}})\! = \!(1,\bar{1})$ [$({j},{{\bar{j}}})\! = \!(2,%
\bar{2})$] to the two energy levels that cross $\varepsilon\! = \!0$ at $%
\phi> \pi$ [at $\phi<\pi$]. Notice that as $\alpha$ is increased the
position of the crossing for $(2,\bar{2})$ at $\varepsilon\! = \! 0$ shifts
in phase to the right whereas for levels $(1,\bar{1})$ the crossing position
remains unchanged. This is interpreted using the WKB picture . We associate levels $(1,\bar{1})$ to states in
which electrons and holes do not go through avoided crossings in the
transport through the QPC. Therefore, in these states the spin remains along
the $y$-direction: i.e., $\eta_{\mathrm{1}}\! \approx \! +1$, for the
Andreev state associated to $|\mathrm{A}\rangle$ [see Fig.~\ref{FG:f4}(A)]
and its $\mathcal{T}$-partner.

On the other hand, the Andreev states corresponding to levels $(2,\bar{2})$
have their spin not fully aligned to the $y$-direction as electrons and holes are affected
by the avoided crossings for nonzero $\alpha$. In the strong SOC limit we
associate them to the Andreev state $|\mathrm{B}\rangle$ [see Fig.~\ref
{FG:f4}(B)] and its $\mathcal{T}$-partner. As the probability to follow the
avoided crossing increases with $\alpha$ [see for example the 2-channels
model in Eq.\eqref{EQ:Tres}] the wavefunction in the region away from the
barrier is changed from being dominated by the first transverse mode (at $%
\alpha\! \approx \!0$ holding $\eta_{\mathrm{2}}\! = \!-1$) to being
dominated by the second transverse mode with opposite spin along the $y$%
-direction (at very strong SOC such that $\eta_{\mathrm{2}}\! \approx \! 1$).

For both types of states we take the slopes of the Andreev levels $%
\lambda_{j}$ and from Eq.\eqref{EQ:slope} compute the values $\delta_{j}$.
Notice that we have sorted out the $\mathcal{T}$-partners (here ${j}\! =
\!1,2$) by increasing ${j}$ as $\lambda_{j}$ decreases, i.e., from the
fastest to the slowest Andreev state. Then we introduce the observed shifts $%
\phi_{{j}}^Z$ (see Fig.~\ref{FG:f7}(Lc)) in Eq.\eqref{EQ:shiftPhi2} to
compute the parameters $\eta_{j}$. The results are presented in Table \ref
{TBjj:vsSOC} and Fig.~\ref{FG:f7}(Lb) where we plot the values $\eta_{j}$.

In Fig.~\ref{FG:f7}(Lb) we see that, as expected, $\eta_1$ is constant as a
function of $\alpha$. We show that $\eta_2$ is well approximated by $2 P_%
\mathrm{av}\! - \! 1$, with $P_\mathrm{av}$ the average polarization of the
QPC in the $\varepsilon\in(-\Delta_0,\Delta_0)$ energy window. This
quantity is a crude estimate for the spin projection of the Andreev state
that suffers the avoided crossing for the case in which the total length of
the junction is much larger than the zone of the barrier in which the spin
gets mixed. In this limit one can assume that the spin of the Andreev state
is dominated by their spin properties away from the barrier, these being the
ones which in the normal device dictate the value of the polarization. More
specifically, we assume that $\eta_2\! \approx \! S_2$, with $S_2$
extracted from the polarization of the normal device. We assume $P_\mathrm{av}\smeq(1+S_2)/2$: the factor $2$ in the denominator accounts for the total number of transmitted modes within the first conductance plateau and the
numerator is the sum of $1$ (i.e., full spin $up$ due to the
electron never leaving the $(1,+)$ mode) with the contribution $S_2$
given by the specific superposition in modes $(1,-)$ and $(2,+)$ arriving to lead $R$ due to the SOC-induced avoided crossing (see 2-channels model
in Sec.\ref{SC:normal}).

\begin{table}[t]
\caption{Dependence with $ \alpha$ of parameters $ \delta_{j}$
and $ \eta_{j}$, ${j}\! = \! 1,2$. The JJ has the quantum point
contact $QPC_1$, the CPRs and Andreev levels are those shown in Fig.~\ref
{FG:f7} (left panels).}
\label{TBjj:vsSOC}\centering
\begin{tabular}{c|llllllll}
\hline\hline
$\alpha$[meVnm]$\rightarrow$ & 0 & 10 & 20 & 30 & 40 & 50 & 60 & 70 \\ \hline
$\delta_1$ & 1.43 & 1.43 & 1.41 & 1.40 & 1.42 & 1.45 & 1.49 & 1.52 \\
$\eta_1$ & 1. & 0.99 & 0.99 & 0.99 & 0.99 & 0.98 & 0.98 & 0.98 \\ \hline
$\delta_2$ & 1.43 & 1.43 & 1.41 & 1.40 & 1.39 & 1.37 & 1.35 & 1.34 \\
$\eta_2$ & -1. & -0.90 & -0.77 & -0.44 & -0.03 & 0.34 & 0.59 & 0.74 \\
\hline\hline
\end{tabular}
\vspace{-0.3cm}
\end{table}

\subsubsection{Asymmetric CPRs}

So far we have shown results for one open channel in the QPC. In the weak-$%
B_y$ limit we have shown that a current appears for zero phase difference.
This is due to different shifts suffered by each of the two $\mathcal{T}$%
-partners of Andreev states. As we discuss in Sec.\ref{SC:SymCPR} here the
current-phase relation does not need to be symmetric and thus the critical
current can depend on the direction of the current flow, i.e., $I^{+}_c \!
\neq \! I^{-}_c$. We find that the latter effect becomes significant when
the barrier is tuned to allow the transmission of a few (more than one) open
channels. This is shown in Fig.~\ref{FG:f7}(Rf) for $N_{oc}\! = \! 2$. For
convenience we number the $\mathcal{T}$-couples ${j}\! = \!1,2,...,2 N_{oc}$
from faster to slower Andreev states (decreasing slope of their dependance on $\phi$). In the case of Fig.~%
\ref{FG:f7}(Rd) and Fig.~\ref{FG:f7}(Re) we show how four of those couples ($%
N_{oc}\! = \!2$) rearrange due to a finite magnetic-field applied along the
direction of polarization. In Table \ref{TBjj:Eqtb1} we present the
parameters $\delta_{j}$ and $\eta_{j}$ obtained from the simulations.
\begin{table}[t]
\caption{Values of $ \delta_{j}$ and $ \eta_{j}$ for two open
channels in the QPCs for different junction lengths when $ \alpha\! =
\! 40$meVnm.}
\label{TBjj:Eqtb1}\centering
\begin{tabular}{cl|ll|ll|ll|ll}
\hline\hline
$V_g$ [meV] & ${L_N}$[nm] & $\delta_1$ & $\eta_1$ & $\delta_2$ & $\eta_2$ & $%
\delta_3$ & $\eta_3$ & $\delta_4$ & $\eta_4$ \\ \hline
$\mathrm{QPC}_1$ & 600 & 1.39 & 0.95 & 1.49 & 0.50 & 1.63 & -0.31 & 1.82 &
0.18 \\
39.4 & 900 & 1.27 & 0.97 & 1.35 & 0.55 & 1.44 & -0.41 & 1.58 & 0.27 \\
& 1200 & 1.21 & 0.99 & 1.27 & 0.58 & 1.33 & -0.45 & 1.47 & 0.36 \\ \hline
$\mathrm{QPC}_2$ & 600 & 1.53 & 0.97 & 1.60 & -0.84 & 1.70 & 0.87 & 1.90 &
0.72 \\
46.3 & 900 & 1.40 & 0.97 & 1.45 & -0.85 & 1.52 & 0.89 & 1.66 & 0.77 \\
& 1200 & 1.31 & 0.99 & 1.36 & -0.86 & 1.41 & 0.90 & 1.52 & 0.79 \\
\hline\hline
\end{tabular}
\vspace{-0.3cm}
\end{table}

As expected the fastest $\mathcal{T}$-pair (${j}\! = \!1$) has $\eta_1\!
\approx \! 1$ due to the fact that they are associated with Andreev states
in the outermost branches of $(1,+)$ and $(1,-)$, i.e., those unaffected by
avoided crossings. The remaining three $\mathcal{T}$-pairs are affected by
the avoided crossings in the transport through the barrier. Each of them
also sees a different effective length ($\delta_{j} {L_N}$) in the passage
through the barrier. This combination of different slopes and different spin
properties ($\eta_{j}$) translates, given a finite magnetic field along the $%
y$-direction, into an asymmetric CPR. The asymmetry is difficult to detect
in the same device for the $N_{oc}\! = \! 1$ case (at a higher gate
voltage). In Fig.~\ref{FG:f7}(Ra) and Fig.~\ref{FG:f7}(Rb) we show that the
rearrangement of Andreev levels for $N_{oc}\! = \! 1$ is symmetrical
(with respect to a new phase value different from $\pi$) except for the small
difference in slopes between the two $\mathcal{T}$-pairs; the latter
difference does not produce a significant asymmetry in the total CPR shown
in Fig.~\ref{FG:f7}(Rc).

The significant difference of the two critical currents, $|I^{+}_c\! - \!
I^{-}_c|/(I^{+}_c\! + \! I^{-}_c)$, also gets reduced in the opposite limit,
when $N_{oc}\gg 1$. In this case the full current is the sum of the
contributions of the $2 N_{oc}$ $\mathcal{T}$-pairs
and balanced spin-properties are dominant since a smaller fraction of the transmitted channels undergo an avoided crossing transition. This can be understood, for example, by taking the height of the
barrier half of what is considered in Fig.~\ref{FG:f1}. Then, the electrons in
channels $(1,-)$ traveling from left to right would not be affected by the
avoided crossings with mode $(2,+)$ and thus both modes $(1,\sigma)$ would
be transmitted. Assuming that the third transverse mode can now transmit,
only the SOC mixing between modes $(2,\sigma)$ and modes $(3,\bar{\sigma})$
contributes to spin filtering effect. This is the reason why
the polarization, and thus the anomalous Josephson effects we study here, is smaller when many channels are allowed to transmit through the barrier.

\subsection{Anomalous Josephson effects enhanced by \newline
barriers with SOC: non-ideal interfaces}

\label{SC:nontransp} In the previous section the analysis was deliberately
made to elucidate the effects introduced by the barrier. Therefore we
have idealized the S-N junctions in order to minimize normal reflections at
the S-N interface. In those simulations the matching between 2DEG and
superconductors was very good despite the fact that the latter have $\alpha\!
= \! 0$ and a smaller g-factor. Here we change the
superconductor system, as described in Sec.\ref{SC:Model},\cite
{ReynosoUB08prl} with the goal of incorporating the mismatch of Fermi
velocities between the 2DEG and the S contacts to simulate more realistic
experimental devices.\cite{Takayanagi95,*Takayanagi95prb} In the following
we show that anomalous effects in the CPR due to the polarizing effects are
present both for realistic junctions based on quantum point contacts and for
wide barriers.

\begin{figure}[b]
\begin{center}
\includegraphics[clip,width=0.48\textwidth]{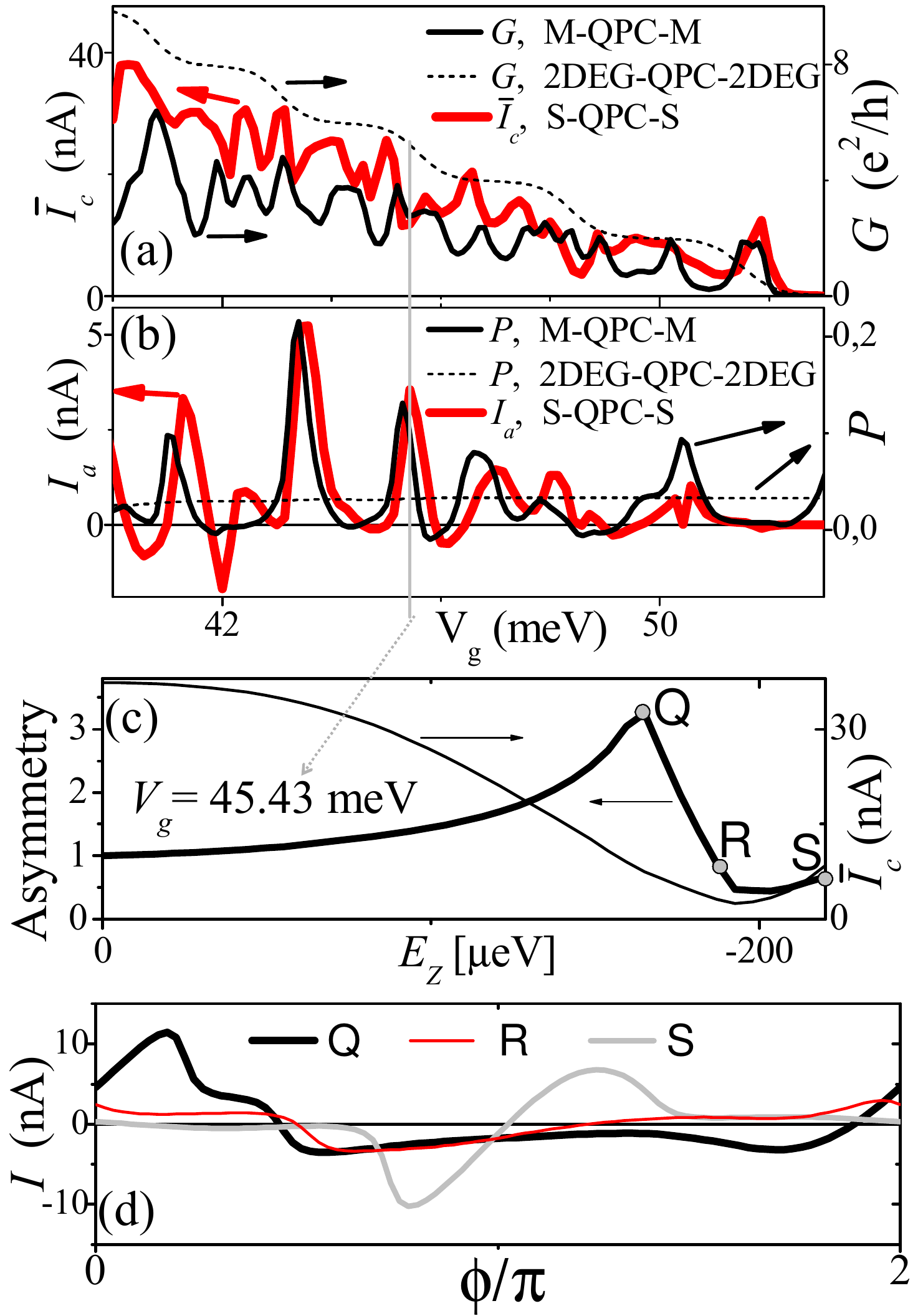}
\end{center}\vspace{-0.5cm}
\caption{Results for a Josephson junction, S-QPC-S, based in QPC$_2$ with
realistic 2DEG-S junctions: an In-based 2DEG and Nb superconductors are
simulated (parameters given in Fig.~\ref{FG:f2}(b)). We take $ \alpha%
\! = \! 5$meVnm, ${L_N}\! = \! 1.2 \mu$m; in (a) and (b) for the JJ
results the Zeeman energy in the 2DEG is $E_Z\! = \!-150  \mu$eV. We
plot the dependence with the gate-voltage, $V_g$, of: (a) the average
critical current, $\bar{I}_c$, and (b) the anomalous Josephson current $I_a$%
. In (a) [(b)] we include the conductance $G$ [the polarization $P$] as a
function of $V_g$ for the same device between normal contacts (dashed line, 2DEG-QPC-2DEG junction) such that the leads do not generate
backscattering, and the leads are
metallic (solid black line, M-QPC-M junction) we take the parameters of Nb with $\Delta_0\! = \! 0$. We observe correlation between $\bar{I}_c$
(${I}_a$) for the S-QPC-S and $G$ ($P$) for the M-QPC-M. (c) We fix $V_g$ at a
peak for $I_a$ and present, as a function of $E_Z$, $\bar{I}_c$ and the
critical current asymmetry $I_c^+/I_c^-$. For such value of $V_g$ the
apartment from the $I_c^+\! = \! I_c^-$ situation is strong and controllable
via magnetic field (and gate voltage as shown in Ref.~\onlinecite{ReynosoUB08prl}). (d) We plot the full current-phase relations
for some situations labeled in (c). }
\label{FG:f8}
\end{figure}

\subsubsection{Quantum point contacts}

The drastic effect of more realistic 2DEG-contact interfaces was discussed
in Sec.\ref{SC:normalRealistic} for the transport properties of QPCs when
the leads were taken to be metallic, i.e. $\Delta_0\! = \!0$. As shown in
Fig.~\ref{FG:f2}---for a ratio between Fermi velocities, $v_%
\mathrm{F}^{\mathrm{2DEG}}/v_\mathrm{F}^{\mathrm{Nb}}\! \approx \!2.25$%
---the conductance through the QPC shows a resonant-like behavior as a
function of gate-voltage. In particular we have shown that, due to quantum
interference effects, the polarization peaks can be larger than the
polarization developed for the case where the 2DEG-contacts interfaces do
not generate backscattering.

Figure \ref{FG:f8} show results derived from the current-phase relation in a
Josephson junction containing a QPC and realistic 2DEG-S interfaces. Due to
the magnetic field $B_y$, the CPR for the system has interesting properties
that can be tuned by changing the gate voltage which controls the barrier
height. For convenience we define the average critical current (see Fig.~\ref
{FG:f8}(a))
\begin{equation}
\bar{I}_c \equiv \frac{1}{2}\left({I}_c^++{I}_c^- \right)\,,
\end{equation}
the anomalous current (or zero phase difference current, see Fig.~\ref{FG:f8}%
(b))
\begin{equation}
I_a \equiv I(\phi\! = \!0)\,,
\end{equation}
and the asymmetry of the CPR (see the dependence with $E_Z$ in Fig.~\ref
{FG:f8}(c))
\begin{equation}
\mathrm{Asymmetry} \equiv \frac{{I}_c^+}{{I}_c^-}\,.
\end{equation}
The underlying resonant tunneling physics produce an abrupt dependence with
the gate voltage of the quantities of interest. This is different from the
idealized S-N junctions case of Sec.\ref{SC:ResultsTR}, for which the
results do not change with $V_g$ providing the number of open channels in
the QPC remains the same. The dependence with $V_g$ shows resonant peaks in
both $\bar{I}_c$ and $I_a$. Fig.~\ref{FG:f8}(d) presents some examples of
CPRs that show a striking asymmetry. In these cases, the fact that the CPRs
fulfill the zero area condition of Eq.\eqref{EQ:zeroArea} is not obvious.

Indeed, because of the strong normal reflection at the interfaces, the
system can be interpreted as a S-quantum dot-S Josephson junction and thus
the resonant like-shapes in the critical current are expected.\cite
{Furusaki91,*Furusaki92} This is the so-called resonant tunneling
supercurrent transistor regime for samples in which the charging energy of
the ``dot" is small and thus Coulomb blockade is not observed.\cite
{Kuhn01,BeenakkerResTunSNS,*Wendin1996} In the absence of a magnetic field the
states arising in the quantum dot appear as time reversal pairs. As for the
case of Sec.\ref{SC:ResultsTR}, for JJs based on normal quantum dots with
spin-orbit coupling, no asymmetric CPRs or current for zero phase difference
are expected at zero field.\cite{BeriBB08}

A correlation between the normal transport and the superconducting transport
is observed in the shapes of $\bar{I}_c$ and $G$ in Fig.~\ref{FG:f8}(a).
This is understood generically, for Josephson transistor devices, because
the critical current is dominated by the contribution of the highest
quasiparticle level (i.e., closest to $\ve\smeq0$).\cite{Kuhn01} Then the critical current is maximized at
the same values of $V_g$ that maximize $G$ in the M-QPC-M device:
constructive interference favors transmission of quasiparticles and Andreev
bound states acquires strong dependence with the phase $\phi$ for $%
\varepsilon\! \approx \! 0$. Moreover, in Fig.~\ref{FG:f8}(b) we find
a strong correlation between $P$ in the normal device and the anomalous current $I_a$
in the associated Josephson junction.

Here the fact that the transport is conditioned by normal backscattering at
the interfaces and interference effects does not invalidate the symmetry
arguments discussed in Sec.\ref{SC:ResultsTR} for the existence of Andreev
states $\mathcal{T}-$partners at zero magnetic field. Indeed, the Andreev
states at a resonant condition are $\mathcal{T}-$partners and it is natural to expect that their spins (see Eq.\eqref{EQ:spinTQpairs}) are correlated with the
polarization found in the corresponding normal junction. At the resonant value of $V_g$
their associated Andreev levels cross each other with opposite slopes at $%
(\varepsilon,\phi)\! = \!(0,\pi)$.\cite{BeenakkerResTunSNS,*Wendin1996} In a
zero SOC system Andreev states are spin degenerate and thus there are two $%
\mathcal{T}-$pairs resonating at the same $V_g$. Here on the other hand, as
pointed out in Sec.\ref{SC:normal}, the two scattering states contributing
to a given transverse mode have different velocities. In the Fabry-Perot
geometry this leads to resonances at different values of $V_g$. In summary,
the multichannel physics, the symmetric sample geometry, and the SOC make a particular spin projection,
$up$ along the $y$-direction in this case, to be favored in the normal transport
from left to right ($P_{L\rightarrow R} >0$). The opposite projection is
favored in the right to left normal transport. The Fabry-Perot effects
enhances the polarizing effect for specific values of $V_g$. In the
associated Josephson junction, when a weak $B_y$ breaks the spin degeneracy, this leads to a nontrivial rearrangements of the Andreev levels
resulting in the nonzero $I_a$ and/or asymmetric CPRs, ${I}_c^+\! \neq \! {{I%
}_c^-}$.

It is worth mentioning that the biggest asymmetry in the CPR occurs due to the magnetic field for the cases that at zero field present a few off-resonance $\mathcal{T}-$pairs (i.e., pairs whose energies are below the Fermi energy for any value of the phase difference) and a single resonant  $\mathcal{T}-$pair (its energy reaches $\ve\smeq 0$ for $\phi\smeq\pi$). In this situation, the dispersion of the resonant Andreev state is close to the one corresponding to the ballistic case while the off-resonant ones are more sinusoidal-like, typical of a low transmission channel (results not shown). Therefore, the shift induced by the $B$ field are very different in each case, leading to the observed asymmetry when the corresponding  contributions to the CPR are added up. This becomes strongly dependent on $V_g$ allowing, at a fixed
$B$, to use slight changes in the gate voltage as a control parameter.\footnote{As shown in Fig.4 of Ref.~\onlinecite{ReynosoUB08prl}.} The ${I}_c^+\!
\neq \!{{I}_c^-}$ condition is generically observed as long as few open
channels are open in the QPC and the modes affected by SOC mixing contribute
a significative fraction of the total current. We observe that the asymmetry
grows with the magnetic field as shown in Fig.~\ref{FG:f8}(c) up to a value
in which it jumps from ${I}_c^+> {{I}_c^-}$ to ${I}_c^+< {{I}_c^-}$. This
dependence is also observed at fixed $B_y$ as a function of the junction
length---in such a case one has to change $V_g$ with ${L_N}$ in order to
follow the resonant condition. The qualitative interchangeability of ${L_N}$
and $B_y$, as in S-F-S junctions, follows because both quantities produce an
enhancement of the underlying phase-shifts of Andreev levels (see Sec.\ref
{SC:phaseShift}).

\begin{figure}[!b]
\begin{center}
\includegraphics[clip,width=0.48\textwidth]{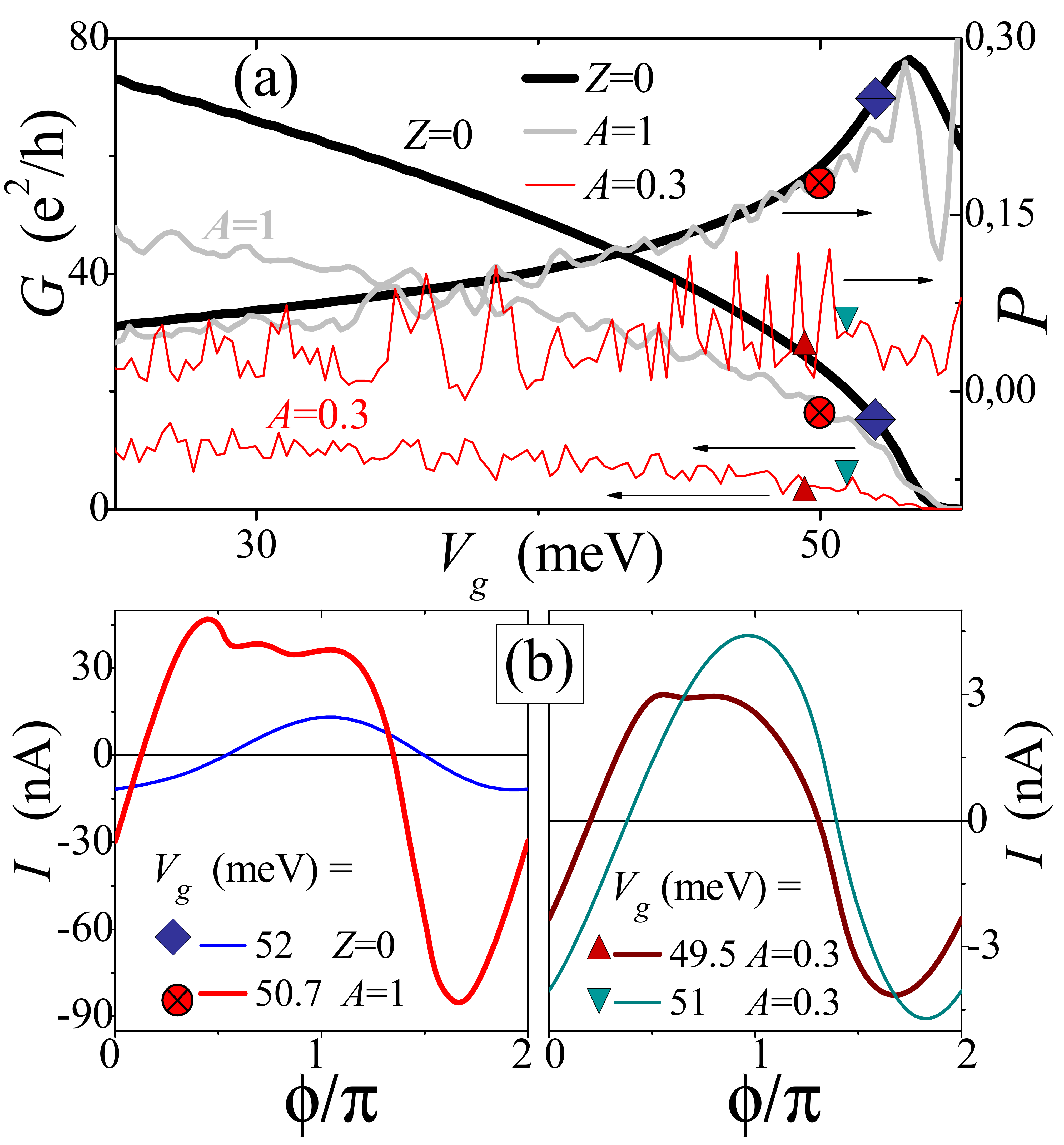}
\end{center} \vspace{-0.5cm}
\caption{Results for a junction with potential $V_{\mathrm{wide}}(x,y)$ with
$L_b\! = \!360$nm, i.e., a plain barrier (B) without a lateral constriction.
We take $ \alpha\! = \!20$meVnm and the 2DEG and superconductor
parameters given in Fig.~\ref{FG:f2} and Fig.~\ref{FG:f8}. (a) Spin
polarizing properties of the barrier when the leads are normal, we show the
dependence on gate voltage $V_g$ for the conductance $G$ and the
polarization $P$. Case $Z\! = \! 0$ is for a 2DEG-B-2DEG junction, i.e., the
interfaces are ideal as the leads are the same In-based 2DEG. The other two
cases are for the M-B-M junction, the leads are metallic Nb and a there is
mismatch of Fermi velocities and effective masses; moreover, away from $A\!
= \! 1$, see Eq.\eqref{EQ:tNtS_A}, the transparency of the junction is
decreased. (b) The normal systems of (a) are turned into Josephson junctions
due to a superconducting pairing $\Delta_0\! = \! 1.5$meV in the contacts
and we also include a Zeeman field along the $y$-axis with energy $E_Z\! =
\!-150 \mu$eV. Current-phase relations showing both current for $%
 \phi\! = \!0$ and/or CPR asymmetry $I_c^+\! \neq \! I_c^-$, as those
appearing in QPCs, can also be obtained for wide barriers.}
\label{FG:f9}
\end{figure}

\subsubsection{Wide barriers}

In Sec.\ref{SC:normal} we have presented the polarizing properties of
QPCs.\cite{EtoHK05} The WKB picture, in Fig.~%
\ref{FG:f1}, shows therefore transverse modes well separated in energy by
virtue of the transversal quantization induced by the lateral constriction
of the QPC. The polarizing properties, however, are also expected for very
wide barriers. The physical mechanism is conceptually the same, i.e., a
smaller transverse modes quantization energy does not avoid the SOC-induced
mode mixing leading to polarization in the transport through an adiabatic
barrier. This nontrivial spin transport shares the same time-reversal
properties as in the discussed QPC. Nonzero polarization has been derived by
Silvestrov and Mishchenko in Ref.~\onlinecite{SilvestrovM05} even for the
case of a very wide system with transversal periodic conditions. As the
barrier (B) becomes smaller the polarization decreases in agreement with the
observations in QPCs. Here we present results for the polarization in a wide
barrier and show that these devices also lead to the same CPR phenomena in
Josephson junctions.

Figure \ref{FG:f9} shows results for the wide barrier potential $V_{\mathrm{%
wide}}(x,y)$ given in Appendix \ref{AP:A}. The barrier has no dependence
with $y$ and thus lateral confinement is determined by the stripe width $W_y$
chosen in the simulation. We take $W_y\! = \! 630$nm, which means $51$ open
channels in the stripe ($V_g\! = \!0$) for our choice of the Fermi
energy. We denote with $Z\! = \!0$  to the case of a 2DEG-B-2DEG junction,
i.e., the leads are essentially the same material as the central region
except from the reduced g-factor and the zero spin-orbit coupling--recall that $Z$ is the amplitude of the barriers at the
S-N junctions in the BTK approach.\cite{BlonderTK82} Then we
study the case in which the leads are metallic, i.e., a M-B-M junction. In
the latter case we use the parameter $A$ presented in Eq.\eqref{EQ:tNtS_A},
to reduce the transparency of the junction away from the ideal condition ($%
A\! = \! 1$).

In Fig.~\ref{FG:f9}(a) we shown the normal transport properties of the barrier, the
conductance $G$ and the polarization $P$, that we use to characterize it. In the $Z\! = \!0$ case, despite
the fact that the conductance plateaux become undefined, the polarization is
nonzero. In the case of metallic leads with high transparency, $A\! = \!1$, the
$G$ and $P$ acquire some structure but without showing marked resonances as in
the QPC case. This can be understood from the fact that many channels with
similar velocity are available and thus neighboring resonances overlap. The
total contribution is similar to the $Z\! = \!0$ case except for a reduction
of the conductance. When we take $A\! = \!0.3$,  the transparency of the
junction is decreased and both the conductance and polarization are affected. In
particular the polarization shows peaks as a function of $V_g$.

We transform all these cases into Josephson junctions by taking $%
\Delta_0\! = \!1.5$meV. The current-phase relations in the presence of a finite magnetic field along the $y-$%
direction is shown in Fig.~\ref{FG:f9}(b). We can
easily find, by tuning $V_g$, situations with $I_a\! \neq \! 0$ and
asymmetric CPRs. As it is the case for QPCs, the asymmetry of the CPRs
becomes larger for the realistic junctions having normal scattering at S-N
interfaces. The phenomena is smaller the more modes are allowed to transmit
in the device as it is also the case for the polarization of the associated
normal device.

\section{Conclusions}

\label{SC:Conclusions} We have described how the spin-polarizing properties
of barriers with spin-orbit interaction affect the spin of the Andreev states. First we revisited the normal device polarization mechanism highlighting the
multichannel nature of the transport while stressing that the polarization of the transmitted electrons is due
to the mode mixing induced by  the SOC term proportional to the transversal
linear momentum. As the device is time-reversal-symmetric strong
restrictions are imposed on the spin-resolved transmissions in opposite
directions. At zero magnetic field the properties derived from these
restrictions do not conspire against the formation of Andreev bound states,
this is opposite to the case of ferromagnetic polarizers.

At zero magnetic field a symmetry of the BdG equation,\cite{Liu2010}
including the SOC, links Andreev states corresponding to phase
differences between superconductors $+\phi_x$ and $-\phi_x$, respectively, that have opposite spin and velocity. The symmetry argument is valid for the
junction irrespective of the transparency of the S-N interfaces. Each of these states, that would appear spin-degenerate in the absence of SOC,
are non-degenerate in its presence. More
remarkable, the polarizing barrier makes the two Andreev states traveling in
the same direction for $+\phi_x$ (with slightly different energy due to SOC)
to share similar spin properties. Because of the mentioned symmetry there
are two Andreev states for $-\phi_x$ that travel in the opposite direction
having opposite spin than those for $+\phi_x$, so the system presents a
spin \emph{chirality}. In our system we have shown that the introduction of a
weak magnetic field along the direction in which the Andreev states are spin
chiral is enough to trigger anomalous Josephson effects such as current at
zero phase difference and $I_c^+\! \neq \! I_c^-$. The chirality condition
here is built into the device properties by the presence of the barrier and thus
it can be switched on and off with gate voltage. The condition is similar to
the one required for nonzero anomalous current in the spin-orbit
coupled quantum dot Josephson junction of Ref.~\onlinecite{Zazunov2009}.
As this spin chirality is intrinsic to other helical systems, we expect that a similar effect could occur in those cases too. That is, a Zeeman induced $I_a$ would also appear if the $B$ field is aligned with the helical direction in Josephson junction based on s-wave superconductors.\cite{Krive2005b,Liu2010}

In order to see how the spin-polarizing properties of the barrier
determine the Andreev states and affect the Josephson effect, we have
idealized the S-N interfaces to minimize normal scattering.
The numerical results agree with a simple picture for the transport through the barrier, based on the analysis of the contribution of two almost degenerated Andreev states with the same velocity and spin for a given phase difference.
In this limit a direct translation of Zeeman energy shifts into phase-shifts of Andreev levels can be made because of the linearity of the phase dispersion of the Andreev levels in the
long junction regime. This, combined with the spin chirality of Andreev
states, leads to $I_a\! \neq \! 0$. We have studied the dependence of the
phase-shifts with junction length, magnetic field and SOC strength.
We also showed that  asymmetric CPRs are possible when the barrier allows more than one open channel.

We also presented results for QPC based
junctions with realistic S-N interfaces. The asymmetry of the
CPRs becomes more marked than in the transparent junction case. This is
related to the Fabry-Perot like transport that produces peaks in the
conductance and polarization of the associated normal device (taking $%
\Delta_0\! = \! 0$ in the leads).

The phenomena described here are to be expected in
barriers as long as the device behaves as a polarizer---due to the mode mixing induced by the SOC. Then, few open channels are preferred. We have shown that anomalous effects are also observable in wide barriers even in cases in which the transparency of the S-N interfaces are not high. Therefore for smooth barriers with few open channels in quantum wires the mentioned effects should be observed as long as the spin-orbit coupling is 2-dimensional, i.e., it must have terms both proportional to the longitudinal momentum and to the transversal momentum. As for quantum wires constriction-type barriers are difficult to implement, the latter is an important result: apart from the practical possibilities arising from the supercurrent rectifier and the phase shifted CPR, the experimental observation of the effects would provide a signature of the presence of a SOC term proportional to the transverse momentum. We believe this is valuable information as SOC is one of the fundamental ingredients required in quantum wires for accessing the topological superconducting phase, and knowledge of all existing SOC terms is desirable.

 \acknowledgments We acknowledge useful discussions with K. Flensberg, S. De Franceschi, L.
Kouwenhoven and A. Zazunov. AAR acknowledges support from the Australian
Research Council Centre of Excellence scheme CE110001013 and from ARO/IARPA
project W911NF-10-1-0330. AAR is grateful for the hospitality of the
Institut N\'eel. GU and CAB acknowledge financial support from PICTs
2008-2236 and Bicentenario 2010-1060 from ANPCyT and PIP 11220080101821 from
CONICET, Argentina. D.F. and M. A. acknowledge support from ECOS-SECyT
collaboration program A06E03 and PICS 05755 from CNRS.

\appendix
\section{}
\label{AP:A}

\subsection{Potential Barrier}

\label{AP:barrier} To describe the quantum point contact we use the realistic potential
given in Ref.~\onlinecite{Ferrybook} for a split-gate defined QPC,
\begin{eqnarray}  \label{EQ:potential}
V_{QPC}(x,y)&=&\frac{V_g \Upsilon(x-x_c,y-y_c)}{\Upsilon(0,0)}\,, \\
\Upsilon(x,y)&=& f\left(\frac{2 x - L_b}{2 z},\frac{2 y + W_b}{2 z}%
\right)-f\left(\frac{2 x + L_b}{2 z},\frac{2 y + W_b}{2 z}\right)  \notag \\
&& +f\left(\frac{2 x - L_b}{2 z},\frac{W_b-2 y}{2 z}\right)-f\left(\frac{2 x
+ L_b}{2 z},\frac{W_b-2 y}{2 z}\right),  \notag \\
f(u,v)&=&\frac{\pi}{2}-\arctan u -\arctan v +\arctan \frac{u v}{\sqrt{1+u^2+
v^2}} \,,  \notag
\end{eqnarray}
where $z$ is the distance between the 2DEG and the plane of the gates, $L_b$
is the length of the gate, $W_b$ is the width of the constriction, and $%
V_{g} $ is the value of the potential at the saddle point located at $%
(x_c,y_c)$.
%

In the case of barriers without the lateral constriction, we use a modified version
of the potential given in Ref.~\onlinecite{AndoQPC1991},
\begin{equation}  \label{EQ:Ando}
V_{\mathrm{wide}}(x,y)=
\begin{cases}
\frac{V_{g}}{2}\left[1\!+\!\cos \left(\frac{\pi (x_{}-x_c)}{L_b}\right)%
\right] & \text{for~}-L_b<x-x_c<L_b \\
0 & \text{otherwise}
\end{cases}
,
\end{equation}
where $2 L_b$ is the barrier length and at the center of the barrier, $x_c$,
the barrier achieves its maximum value $V_g$.

\subsection{Tight binding model}

\label{AP:TB} Following standard finite-difference procedures for a square
lattice with spacing $a_0$, we write the total Hamiltonian as
\begin{equation}
\hat{H} \! = \! \hat{H}_{Z}\! + \! \hat{H}_{N}\! + \! \hat{H}_{{L}}\! + \!
\hat{H}_{{R}}\! + \!\hat{H}_{N,{L}}\! + \!\hat{H}_{N,{R}}\,.
\end{equation}
For convenience we have used the same notation as the continuos
Hamiltonians given in Sec.\ref{SC:Model} and we have introduced the
Hamiltonians $\hat{H}_{N,\gamma}$ connecting the normal region and the lead $%
\gamma\! = \!\{{L},{R}\}$. The Hamiltonian \eqref{EQ:hc} for the
normal region becomes,
\begin{eqnarray}
\hat{H}_N &=& \sum_{{\bm r}\in \mathrm{N},\sigma}{(4t_N+V({\bm r})-\mu)\hat{c%
}_{{\bm r},\sigma}^{\dag}} \hat{c}_{{\bm r},\sigma}^{}- \left[\sum_{{\bm r}%
\in \mathrm{N},\sigma}t_N\hat{c}_{{\bm r},\sigma}^{\dag} \hat{c}^{}_{{\bm r}
+ a_0{\hat{\mathbf{x}}},\sigma} \right.  \notag \\
&& \left.  \sum_{{\bm r}\in \mathrm{N},\sigma} t_N \hat{c}_{{\bm r}%
,\sigma}^{\dag} \hat{c}^{}_{{\bm r} + a_0{\hat{\mathbf{y}}},\sigma} -\sum_{{%
\bm r}\in \mathrm{N},\sigma} \lambda_{x,\sigma\bar{\sigma}} \hat{c}_{{\bm r}%
,\sigma}^{\dag}\hat{c}^{}_{{\bm r} + a_0{\hat{\mathbf{x}}},\bar{\sigma}}
\right.  \notag \\
&& \left. -\sum_{{\bm r}\in\mathrm{N},\sigma} \lambda_{y,\sigma\bar{\sigma}}
\hat{c}_{{\bm r},\sigma}^{\dag}\hat{c}^{}_{{\bm r} + a_0{\hat{\mathbf{y}}},%
\bar{\sigma}} +h.c. \right],  \label{EQ:HTBnorm}
\end{eqnarray}
where $\hat{c}_{{\bm r},\sigma}^{}$ ($\hat{c}_{{\bm r},\sigma}^\dag$) refers
to the annihilation (creation) operator for an electron in the 2DEG at
position ${\bm r}$ with spin $\sigma_z\! = \!\sigma\! =
\!\{\uparrow,\downarrow\}$. The hopping is $t_N\! = \!\hbar
^{2}/2m^{*}a_{0}^{2}$. The summations are taken over all lattice sites in
the normal region, N; terms with operators $\hat{c}_{{\bm r},\sigma }$ and $%
\hat{c}_{{\bm r},\sigma }^{\dag}$ with ${\bm r}$ outside such region are
taken to be zero. The Rashba spin-orbit coupling parameters are related to the
Rashba strength in the continuos model as
\begin{eqnarray}
\lambda_{x,\uparrow\downarrow}\! = \! \frac{\alpha}{2a_0}~,~~\lambda_{x,%
\downarrow\uparrow}\! = \! -\frac{\alpha}{2a_0} ~,  \notag \\
\lambda_{y,\uparrow\downarrow}\! = \! -\mathrm{i}\frac{\alpha}{2a_0} ~,~~
\lambda_{y,\downarrow\uparrow}\! = \! -\mathrm{i}\frac{\alpha}{2a_0} ~.
\end{eqnarray}
Notice that time-reversal symmetry is intrinsic to the SOC coupling as we
have $\lambda_{j,\sigma\bar{\sigma}}\! = \!-\left(\lambda_{j,\bar{\sigma}%
\sigma}\right)^*$ with $j\! = \!\{x,y\}$.

The lattice version of the Hamiltonian of the leads \eqref{EQ:Hleads} is
\begin{eqnarray}
\hat{H}_\gamma &=& \sum_{{\bm r}\in R_\gamma,\sigma}{(4t_S + E_S -\mu)\hat{d}%
_{{\bm r},\sigma}^{\dag}} \hat{d}^{}_{{\bm r},\sigma}- \left[\sum_{{\bm r}%
\in R_\gamma,\sigma}t_S\hat{d}_{{\bm r},\sigma}^{\dag} \hat{d}^{}_{{\bm r} +
a_0{\hat{\mathbf{x}}},\sigma} \right.  \notag \\
&&\!\!\!\! \left. + \sum_{{\bm r}\in R_\gamma,\sigma} t_S \hat{d}_{{\bm r}%
,\sigma}^{\dag} \hat{d}^{}_{{\bm r} + a_0{\hat{\mathbf{y}}},\sigma}
+\Delta_0 \sum_{{\bm r}\in R_\gamma,\sigma} \hat{d}_{{\bm r},\uparrow}^{\dag}%
\hat{d}_{{\bm r} ,\downarrow}^{\dag} +h.c. \right]\,,~~~~~~
\end{eqnarray}
where $\gamma\! = \!\{R,L\}$ for the right and left lead, respectively, the
summations are taken over sites belonging to the regions of those leads $%
R_\gamma$ and $\hat{d}_{{\bm r},\sigma}^{}$ ($\hat{d}_{{\bm r},\sigma}^\dag$%
) refers to the annihilation (creation) operator for an electron in the lead
$\gamma$ at position ${\bm r}$ with spin $\sigma_z\! = \!\sigma\! =
\!\{\uparrow,\downarrow\}$. We use the parameters $t_S$ and $E_S$ for
simulating the different types of leads.

When the goal is to minimize the scattering at the leads we use $E_S\! = \! 0$
and $t_S\! = \! t_N$ (and $A\! = \!1$, see Eq.\eqref{EQ:tNtS_A}). On the
other hand, for simulating realistic superconducting leads (or metallic ones
if $\Delta_0\! = \! 0$), we change both $E_S$ and $t_S$
assuring that the lead has many open channels---with the appropiated velocity (dispersion)---in a large energy window around the Fermi energy. The size of the latter is not important in the metallic phase, as the low temperature transport is dominated by the quantum transmission at the Fermi energy. In the superconducting case, it is enough to consider an energy window of a few $\Delta_0$, as the Josephson current in S-N-S junctions is mediated by
Andreev reflection which decays exponentially as a function of energy for $%
|\varepsilon|>\Delta_0$.\cite{BlonderTK82} The contribution to the CPR of
states at energies outside this window is negligible.

In writing the Hamiltonians $\hat{H}_\gamma$ we have chosen a gauge where the superconducting
phases $(\phi_l,\phi_r)=(0,-\phi)$ and  $\phi$ is
accumulated only in the tunneling Hamiltonian between the central 2DEG and
the right superconducting lead.\cite{Cuevas96} We group the lattice sites
according to the coordinate $x$. The set $X_{R}$ ($X_{L}$), having the sites
at $x\! = \! x_{R}$ ($x\! = \! x_{L}$), is the last (first) vertical layer
in the normal region. The intermediate Hamiltonians are
\begin{eqnarray}
\hat{H}_{N, \gamma} &=& \left[-\sum_{{\bm r}\in X_\gamma,\sigma} {t_b
\mathrm{e}^{\mathrm{i} \vartheta_\gamma} \hat{c}_{{\bm r},\sigma}^{\dag}
\hat{d}^{}_{{\bm r} +s_\gamma a_0{\hat{\mathbf{x}}},\sigma}} \right.  \notag
\\
&&\left.+\sum_{{\bm r}\in X_\gamma,\sigma} \frac{\lambda^{{\gamma}}_{\sigma%
\bar{\sigma}}}{2} \mathrm{e}^{\mathrm{i} \vartheta_\gamma} \hat{c}_{{\bm r}%
,\sigma}^{\dag}\hat{d}^{}_{{\bm r} + s_\gamma a_0{\hat{\mathbf{x}}},\bar{%
\sigma}} +h.c. \right], \label{EQ:APtunn}
\end{eqnarray}
with parameters $(s_{L},\lambda^{L}_{\uparrow\downarrow},\lambda^{L}_{%
\downarrow\uparrow},\vartheta_{L})=(-1,\lambda_{x,\downarrow\uparrow}^*,%
\lambda_{x,\uparrow\downarrow}^*,0), $ and $(s_{R},\lambda^{r}_{\uparrow%
\downarrow},\lambda^{r}_{\downarrow\uparrow},\vartheta_{r})=(1,\lambda_{x,%
\uparrow\downarrow},\lambda_{x,\downarrow\uparrow},-\phi/2)$. Notice that
the variation of the Rashba strength---as $\alpha$ is taken to be zero in
the leads and non-zero in the central region---introduces a SOC contribution
to the tunneling Hamiltonian at the interfaces. This follows as a result of
the correct discretization of the SOC term $\frac{\alpha(x)}{\hbar} p_x
\sigma_y$---when $\frac{%
d\alpha(x)}{dx}\! \neq \! 0$, i.e., at $x\! = \! x_{L}$ and $x\! = \! x_{R}$, this term needs to be symmetrized to be hermitian.

Finally, the Zeeman term in the normal region is just
\begin{eqnarray}
\hat{H}_Z &=& \frac{g\mu_B}{2}\left\{\sum_{{\bm r},\sigma}{\
\left(\delta_{\uparrow,\sigma}-\delta_{\downarrow,\sigma}\right) B_z \hat{c}%
_{{\bm r},\sigma}^{\dag}} \hat{c}^{}_{{\bm r},\sigma} \right. \\
&& \left.+ \left[\sum_{{\bm r}}{\left(B_x-\mathrm{i} B_y\right)\hat{c}_{{\bm %
r},\uparrow}^{\dag} \hat{c}^{}_{{\bm r},\downarrow}} +h.c.\right]\right\}\,,
\notag  \label{EQ:zeeman}
\end{eqnarray}
The Zeeman interaction at the leads $\gamma\! = \!\{L,R\}$ have the same
form but in terms of the operators $\hat{d}^{}_{{\bm r},\sigma}$ and $\hat{d}%
^{\dag}_{{\bm r},\sigma}$ with summations taken for ${\bm r}\in R_\gamma$
and $g_N$ replaced by $g_S$.

\subsubsection{Conductance and Polarization}

For the normal case ($\Delta_0\! = \! 0$) we compute the transport properties in
the linear response regime. The conductance follows from
\begin{equation}
G = \frac{e^{2}}{h} \mathrm{Tr}\left[\bm{\Gamma }^{L}\bm{G}^{r} (E_{\text{F}%
})\bm{\Gamma }^{R}\bm{G}^{a} (E_{\text{F}})\right]\,,
\end{equation}
where $\bm{G}^{r(a)}(\varepsilon)$ is the retarded (advanced) matrix
propagator, with elements $\bm{G}_{{\bm r}\sigma ,{\bm r}^{\prime}\sigma
^{\prime }}^{r}(\varepsilon)$ given by the propagator from site ${\bm r}%
^{\prime}$ and spin $\sigma^{\prime}$ to site ${\bm r}$ and spin $\sigma$,
$\bm{\Gamma }_{{\bm r}\sigma ,{\bm r}^{\prime}\sigma ^{\prime }}^{{\gamma}%
}\! = \!{\mathrm{i}}(\bm{\Sigma}_{{\gamma}}^{r}\! - \!\bm{\Sigma}_{{\gamma}%
}^{a})_{{\bm r}\sigma ,{\bm r}^{\prime}\sigma ^{\prime }}$ where $\bm{\Sigma}%
_{{\gamma}}^{r(a)}$ is the retarded (advanced) self-energy due to the $%
\gamma $ lead and $E_{\text{F}}$ is the Fermi energy.

The polarization is
\begin{equation}
P =\frac{1}{G} \sum_{\sigma }(G_{\uparrow \sigma }\!-\!G_{\downarrow \sigma
})\,,  \label{EQ:pola}
\end{equation}
with $G_{\sigma ^{\prime }\sigma }$ the spin-resolved conductances, i.e.,
the contributions due to the electrons that are injected from lead $L$ with
spin $\sigma$ and collected at lead $R$ with spin $\sigma ^{\prime }$.
As the current flow is in the $x$-direction, the relevant polarization occurs for the $y$-axis---recall that the Rashba SOC is proportional to $p_x\sigma_y$. Thus, Eq.\eqref{EQ:pola} refers to the
spin-resolved conductances with $\sigma\! = \!\{\uparrow,\downarrow\}$ along
the $y$-axis. The polarizations along the $x$
and $z$ spin-axis are zero.

\subsubsection{Josephson current}

The Josephson current \eqref{EQ:iphi0}, being proportional to $\left\langle
\frac{\partial \hat{H}}{\partial \phi }\right\rangle$, follows from $\hat{H}%
_{N, r}$ as it is the only term in $\hat{H}$ that depends on $\phi$,
\begin{eqnarray}
I &=&\frac{2e}{\hbar }\mathrm{i}\left(\sum_{\sigma, {\bm r} \epsilon X_{{R}}
}\left[ -\frac{t_{b} }{2}e^{-\mathrm{i}\frac{\phi}{2}}\langle \hat{c}_{{\bm r%
},\sigma }^{\dag }\hat{{d}} _{{\bm r}+a_{0}\hat{x},\sigma }^{}\rangle +\frac{%
t_{b}}{2}e^{\mathrm{i}\frac{\phi}{2}}\langle \hat{{d}}_{{\bm r}+a_{0}\hat{x}%
,\sigma }^{\dag }\hat{c}_{{\bm r},\sigma }^{}\rangle \right] + \right.
\notag \\
&&\sum_{{\bm r}\epsilon X_{{R}} }\left[ \frac{\lambda _{x,\uparrow
\downarrow }}{4}e^{-\mathrm{i}\frac{\phi}{2}}\langle \hat{c}_{{\bm r}%
,\uparrow }^{\dag }\hat{{d}}_{{\bm r}+a_{0}\hat{x},\downarrow }^{}\rangle -%
\frac{\lambda _{x,\uparrow \downarrow }^{\ast }}{4}e^{\mathrm{i}\frac{\phi}{2%
}}\langle \hat{{d}}_{{\bm r}+a_{0}\hat{x},\downarrow }^{\dag }\hat{c}_{{\bm r%
},\uparrow }^{}\rangle \right. +  \notag \\
&&\left.\left. \frac{\lambda _{x,\downarrow \uparrow }}{4}e^{-\mathrm{i}%
\frac{\phi}{2}}\langle \hat{c}_{{\bm r},\downarrow }^{\dag }\hat{{d}}_{{\bm r%
}+a_{0}\hat{x},\uparrow }^{}\rangle -\frac{\lambda _{x,\downarrow \uparrow
}^{\ast }}{4}e^{\mathrm{i}\frac{\phi}{2}}\langle \hat{{d}}_{{\bm r}+a_{0}%
\hat{x},\uparrow }^{\dag }\hat{c}_{{\bm r},\downarrow }^{}\rangle \right]
\right)~.  \label{ECjj:iphi}
\end{eqnarray}
This quantity is computed using normal propagators between the sites in the $%
X_{R}$ layer and sites in the first superconducting layer of the ${R}$ lead. As $%
\Delta_0\! \neq \! 0$ the calculation of the required propagators requires to work in the Nambu space since the equations of motion for normal and anomalous propagators are coupled. We write
all the coupled equations using the standard method of Ref.~\onlinecite{Zubarev60} and obtain the normal propagators $\bm{G}_{{\bm r}%
\sigma,{\bm r}^{\prime}\sigma^{\prime }}^{r(a)}(\varepsilon)$.

\subsubsection{Density of states, spin and current densities}

After solving the equations of motion for the
propagators, all the densities are obtained from,\cite{Zubarev60}
\bese
\begin{eqnarray}  \label{EQ:propaInt}
\langle \hat{\zeta}_{{\bm r}^{\prime},\sigma^{\prime} }^{\dag }\hat{\zeta}_{{%
\bm r},\sigma}^{}\rangle &=& \int_{-\infty}^{\infty}{\rho_{{\bm r}\sigma,{%
\bm r}^{\prime}\sigma^{\prime}}(\varepsilon) f_0(\varepsilon)\,d\varepsilon}%
\,, \\
\rho_{{\bm r}\sigma,{\bm r}^{\prime}\sigma^{\prime}}(\varepsilon) &\equiv&
\frac{\mathrm{i}}{2\pi}\left(\bm{G}_{{\bm r}\sigma,{\bm r}%
^{\prime}\sigma^{\prime }}^{r}(\varepsilon)-\bm{G}_{{\bm r}\sigma,{\bm r}%
^{\prime}\sigma^{\prime }}^{a}(\varepsilon)\right)\,,~~~~
\end{eqnarray}
\eese
where the fermionic operators $\hat{\zeta}$ are type $\hat{c}$, or $\hat{{d}}$ according to the region in which they operate.

When only the integrated quantity is important, for instance when studying the total Josephson current as a function of an external parameter,  the integral in Eq.\eqref{EQ:propaInt} can be computed in the complex plane by using the Residue Theorem. In this case, the number of integration points required for convergence is reduced by a factor of $10$ as we avoid the problem of integrating the chain of Dirac-$\delta$s found in the real axis for $|\varepsilon|<\Delta_0$.

%

\end{document}